\documentclass[sigconf]{acmart}
\usepackage{graphicx}
\usepackage{multirow}

\AtBeginDocument{%
  \providecommand\BibTeX{{%
    \normalfont B\kern-0.5em{\scshape i\kern-0.25em b}\kern-0.8em\TeX}}}

\copyrightyear{2023} 
\acmYear{2023} 
\setcopyright{acmlicensed}\acmConference[CHI '23]{Proceedings of the 2023 CHI Conference on Human Factors in Computing Systems}{April 23--28, 2023}{Hamburg, Germany}
\acmBooktitle{Proceedings of the 2023 CHI Conference on Human Factors in Computing Systems (CHI '23), April 23--28, 2023, Hamburg, Germany}
\acmPrice{15.00}
\acmDOI{10.1145/3544548.3580658}
\acmISBN{978-1-4503-9421-5/23/04}

\begin{document}
\title{A Human-Centered Review of Algorithms in Decision-Making in Higher Education}
%% The "author" command and its associated commands are used to define the authors and their affiliations.
%% Of note is the shared affiliation of the first two authors, and the "authornote" and "authornotemark" commands used to denote shared contribution to the research.
\author{Kelly McConvey}
\orcid{0000-0002-1320-7401}
\affiliation{%
    \institution{University of Toronto}
    \streetaddress{27 King's College Circle}
  \city{Toronto}
  \state{Ontario}
  \country{Canada}
  \postcode{43017-6221}}
\email{kelly.mcconvey@mail.utoronto.ca}

\author{Shion Guha}
\orcid{0000-0003-0073-2378}
\affiliation{%
  \institution{University of Toronto}
  \city{Toronto}
  \state{Ontario}
  \country{Canada}}
\email{shion.guha@utoronto.ca}

\author{Anastasia Kuzminykh}
\orcid{0000-0002-5941-4641}
\affiliation{%
  \institution{University of Toronto}
  \city{Toronto}
  \state{Ontario}
  \country{Canada}}
\email{anastasia.kuzminykh@utoronto.ca}

\begin{abstract}
The use of algorithms for decision-making in higher education is steadily growing, promising cost-savings to institutions and personalized service for students but also raising ethical challenges around surveillance, fairness, and interpretation of data. To address the lack of systematic understanding of how these algorithms are currently designed, we reviewed an extensive corpus of papers proposing algorithms for decision-making in higher education. We categorized them based on input data, computational method, and target outcome, and then investigated the interrelations of these factors with the application of human-centered lenses: theoretical, participatory, or speculative design. We found that the models are trending towards deep learning, and increased use of student personal data and protected attributes, with the target scope expanding towards automated decisions. However, despite the associated decrease in interpretability and explainability, current development predominantly fails to incorporate human-centered lenses. We discuss the challenges with these trends and advocate for a human-centered approach. 
\end{abstract}
%  \Description{Enjoying the baseball game from the third-base seats. Ichiro Suzuki preparing to bat.}
\maketitle
\section{Introduction}
%Growing Use of algorithms
The use of algorithms for decision-making in higher education, and subsequently the use of student data in algorithms, is growing across the globe \cite{yan_trends_2021}. Prior research has found that algorithmic decision-making has the potential to provide considerable cost savings to higher education institutions, with more personalized and just-in-time service for students \cite{marcinkowski_implications_2020}. Indeed, students, faculty, and the administration of higher education institutions face growing challenges; increasing tuition fees and debt levels, along with lower levels of government support, have impacted students and higher education institutions alike \cite{yan_trends_2021}. Neo-liberalism and the rise of knowledge capitalism within higher education have pushed higher education institutions towards a greater emphasis on metrics, accountability, and KPIs \cite{olssen__neoliberalism_2005}. This has led to the growing use of educational data mining, reliance on learning analytics \cite{abdul_jalil_learning_2021} and the use of algorithms for decision-making, predictions, interventions, and personalization \cite{bajpai_big_2017}. 

With this growing trend, both the improper use of student data and the potential for harmful decision-making (predicted and/or automated) within higher education institutions have also risen \cite{marcinkowski_implications_2020,williamson_learning_2021}.
%Higher Ed Challenges
%Pros and cons of Educational Data Mining
In particular, the use of algorithms in education is associated with several ethical challenges, such as student surveillance and privacy, fairness and equity, and interpretation of data \cite{slade_learning_2013}. 
%Previous Research in Higher Ed
The harmful outcomes of ignoring these and similar ethical issues in algorithm design have recently brought forward significant concerns. The SIGCHI community has considered that algorithm design has failed to identify the true target of intervention \cite{barocas_putting_2014}. Abebe et al. \cite{abebe_roles_2020} raise the example of admissions in higher education in that "a computational intervention that aims to equalize offers of college admission across demographic groups might function as a less ambitious substitute for the deeper and more challenging work of improving high school instruction in low-income neighborhoods." Chancellor et al. \cite{chancellor_who_2019} posit that traditional computational research minimizes individuals to simple data points, and there is cause for concern: machine learning can amplify stigma, reproduce stereotypes, increase discriminatory practices, and harm individuals and communities.   

Correspondingly, research is actively exploring different techniques for addressing these ethical issues, including developing frameworks \cite{otoo-arthur_systematic_2019}, and governance strategies \cite{self_governance_2014}, attempting to build more fair models \cite{kusner_counterfactual_2017, lee_evaluation_2020, yu_towards_2020, li_yet_2021}, assessing the need for protected attributes as input data \cite{yu_should_2021}, and incorporating student perspectives \cite{marcinkowski_implications_2020, mcpherson_student_2016}. 

%SIGCHI Problem
As the SIGCHI community continues to pursue research on equity and bias in algorithmic decision-making  \cite{chancellor_who_2019, delgado_uncommon_2022, flugge_algorithmic_2020, kim_human-centered_2021, razi_human-centered_2021, saxena_human-centered_2020, aragon_human-centered_2022, shneiderman_human-centered_2022},
human-centered algorithm design attempts to enable and extend these techniques by incorporating human and social interpretations into the design of algorithmic systems. Specifically, it was suggested \cite{baumer_toward_2017} that theoretical, participatory, and speculative strategies can be employed to center humans in the design process and to bridge the gap between the algorithm developers and the stakeholders who interact with the system or are affected by their decisions.  
With the significance of the social impact of algorithmic decision-making on various high-stakes human domains \cite{chancellor_who_2019, delgado_uncommon_2022, flugge_algorithmic_2020, kim_human-centered_2021, razi_human-centered_2021, saxena_human-centered_2020}, there is concern within the HCI research community that the design process of algorithms fails to consider the potential for harm as a result of the inherent uncertainties of the predictions and limitations of the technology \cite{vaughan_human-centered_2022, razi_human-centered_2021}. Human-centered algorithm design addresses this concern by leveraging human knowledge from the social sciences and incorporating stakeholder perspectives, allowing researchers to better consider the impacts of algorithmic decision-making in the real world or our context \cite{alejandro_jaimes_human-centered_2007}.

However, while research in other domains such as child welfare \cite{saxena_human-centered_2020}, cyberbullying detection \cite{kim_human-centered_2021}, law \cite{delgado_uncommon_2022}, online sexual risk detection \cite{razi_human-centered_2021} and public administration \cite{flugge_algorithmic_2020} has demonstrated both the lack of and the pressing need for human-centered algorithm design \cite{kim_human-centered_2021}, little is known about the application of a human-centered lens in the design of decision-making algorithms in higher education \cite{luzardo_estimation_2014, karypis_improving_2017}.

To address this gap, in this paper, we explore the current trends in the use of computational methods, data types, and target outcomes in the design of algorithms in higher education and analyze the current role and place of human-centered algorithm design approaches in their development. Through a comprehensive review of the existing literature on algorithmic design for higher education (n=62), we collect and qualitatively analyze the models proposed from 2010 to 2021 and demonstrate the existing patterns. 

We show that, first, the model design has trended away from rules-based systems towards neural networks and natural language processing. By nature of the model design, the results have become less explainable and less interpretable and increasingly rely on individual student data such as GPAs, enrolment pathways, and Learning Management System (LMS) activity as input features. We also find that the use of protected classes (age, race, gender, disability status) as input features has grown significantly over the past decade, along with the use of the student or applicant's family data such as household income or parental academic achievement (first-in-family).
At the same time, while the models are increasingly complex and the decisions become increasingly opaque, the algorithm design does not demonstrate the systematic use of human-centered approaches to reflect the necessary student perspective appropriately.

This work contributes to the community by presenting an in-depth account of the current state-of-the-art and trends in algorithmic decision-making in higher education and critically reviewing the algorithms proposed for use in higher education through the application of the human-centered conceptual framework \cite{baumer_toward_2017}. 
%\item An in-depth account of the current state-of-the-art and trends in algorithmic decision-making in higher education. 
Moreover, based on our findings, we identify potential gaps in the existing literature and suggest future research opportunities for developing human-centered algorithms for higher education.
Building upon existing reviews \cite{saxena_human-centered_2020,kim_human-centered_2021,razi_human-centered_2021} on the use of human-centered design in algorithm development in different domains, this work provides a foundation for implementing human-centered approaches in the design and development of algorithms in the context of higher education.

In the remainder of this paper, we first review the existing literature on algorithmic decision-making in higher education and the application of Human-Centered Algorithm Design in other domains. We then describe our data collection and data analysis processes, followed by the results of the analysis of the models proposed in the collected literature corpus. 
Finally, we discuss the critical gaps in the current trends identified through our analysis and propose key opportunities for future research.
\begin{figure}[t]
\centering
\caption{PRISMA Flow Diagram \cite{moher_preferred_2009}}\label{tab:PRISMA Flow Diagram}
\includegraphics[width=8.5cm]{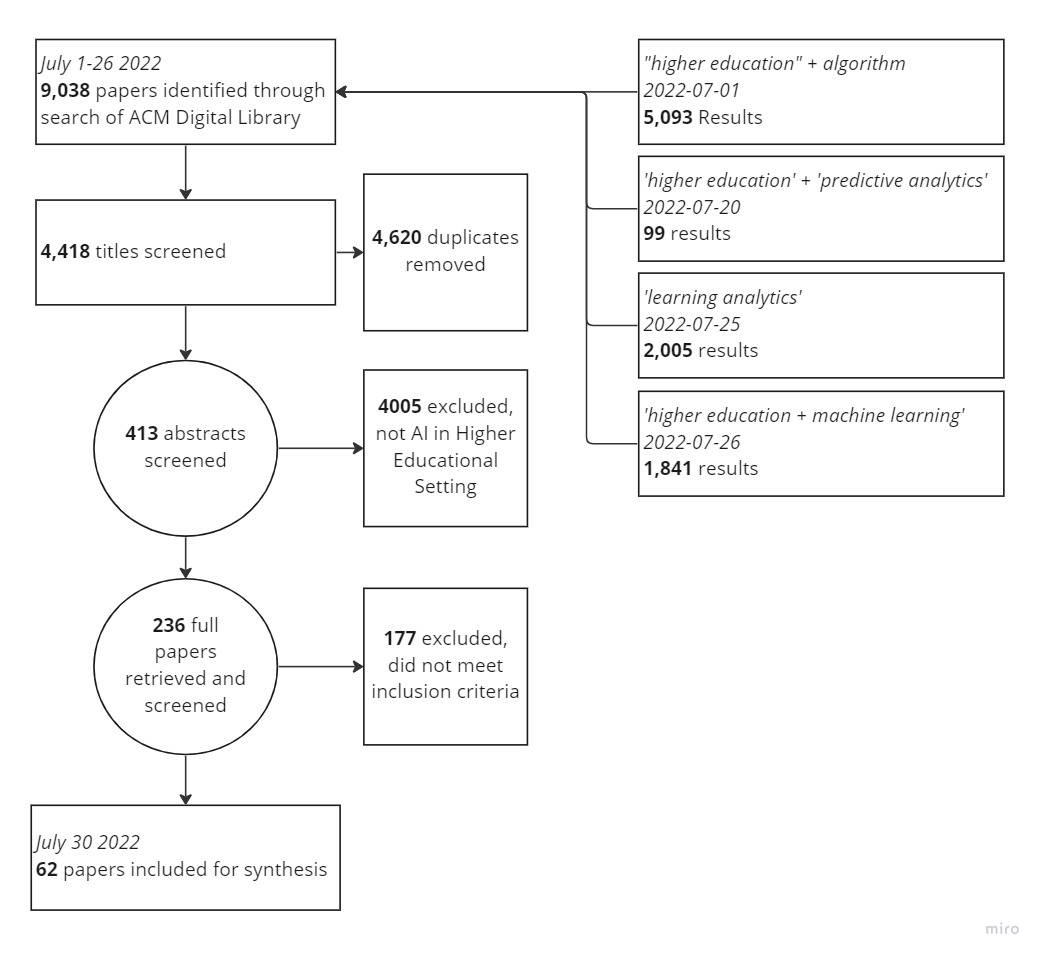}
\end{figure}
\section{Background}
In this section, we provide an overview of the existing literature surveys on the use of algorithms in higher education. We also provide background on the Human-Centered Algorithm Design framework \cite{baumer_toward_2017} used throughout our research and on how it has been used within other domains.
\begin{table*}[]
\centering
\caption{Coding Categories}\label{tab:Coding Categories}
\begin{tabular}{lll}
\hline
\textbf{Computational   Method} & \textbf{Target Outcome} & \textbf{Input Data} \\ \hline
Statistical Methods             & Grade Prediction        & Demographic         \\
Rules-Based                     & Retention               & LMS/Engagement      \\
Machine Learning                & Institutional Planning  & Institutional       \\
Deep Learning                   & Pathway Advising        & Grade/GPA           \\
Natural Language Processing     & Student Services        & Enrollment/Pathways \\
                                & Admissions              & Student Survey      \\
                                & Assessment              & Protected Attribute \\
                                & Engagement                                    \\ \hline
\end{tabular}
\end{table*}
Survey research and systematic literature reviews are important contributions to HCI research, providing insight into what is currently known about the topic at hand, exposing trends and gaps, and identifying opportunities for further research \cite{wobbrock_research_2016}. Previous literature reviews of algorithms in higher education have predominantly focused on model accuracy and performance, rather than fairness through human-centeredness. Nine of the ten reviews that we identified all focused on the same outcome: predicting student success \cite{albreiki_systematic_2021,alyahyan_predicting_2020,campbell_predicting_1996,hellas_predicting_2018, kumar_literature_2017,mouw_prediction_1993,ofori_using_2020,sandra_machine_2021,shahiri_review_2015}. The remaining study \cite{zawacki-richter_systematic_2019} explored various 'AITech' target outcomes. These reviews looked at input data \cite{albreiki_systematic_2021,alyahyan_predicting_2020, campbell_predicting_1996,hellas_predicting_2018, kumar_literature_2017,ofori_using_2020,sandra_machine_2021, shahiri_review_2015}, computational methods \cite{albreiki_systematic_2021,alyahyan_predicting_2020, hellas_predicting_2018, kumar_literature_2017,ofori_using_2020,sandra_machine_2021, shahiri_review_2015,zawacki-richter_systematic_2019} and model performance \cite{kumar_literature_2017,ofori_using_2020, shahiri_review_2015,zawacki-richter_systematic_2019}. Only one paper \cite{hellas_predicting_2018} included any analysis of ethical issues within its reviewed papers; Hellas et al. \cite{hellas_predicting_2018} briefly raise ethical considerations in their discussion. 

\label{ssec:Human-Centered Algorithm Design}
Overwhelmingly, the research on algorithms in higher education has been focused on optimizing the algorithms themselves: the inputs, target outcomes, architecture, and performance. 
The target outcomes aim to make predictions that often impact those humans' lives: what courses they take, what interventions are offered to them, and what programs they are admitted to. Algorithmic performance, though, is a measurement of the functioning of the algorithm and may not align with human and social interpretations of a model's success (does it do and mean what it claims). So while algorithms are developed using staggering amounts of information about humans to then make decisions for humans, it is alarming that humans are so inauspiciously missing from both the design and measurement of their value.

%further describe the issues
To address these issues, Baumer \cite{baumer_toward_2017} recommends three strategies for human-centered algorithm design: theoretical, participatory, and speculative.

\emph{Theoretical Design}, according to Baumer, incorporates behavioral and social science theories into the design of the algorithm. These theories can be used prescriptively to guide algorithm design, informing feature selection, for example, and used descriptively to help us to interpret and evaluate the results of the algorithm.

A \emph{Participatory Design} approach to algorithm design incorporates stakeholders in the design process, the people whom the system will likely impact. In the case of algorithm design and machine learning, this involves connecting the people for whom the algorithm will automate decisions and the end users with the designers, and actively considering the user experience of the system.

\emph{Speculative Design}, according to Baumer, requires an imaginative approach. Researchers must not only consider the existing circumstances but must extrapolate from it what \textit{could} be. For algorithm design, this requires authors to think through the potential impacts and ramifications of the assumptions and values embedded in the ground truth of the model. 

Other domains have demonstrated the feasibility of employing Baumer's human-centered algorithm design framework \cite{baumer_toward_2017}. In a review of algorithms used in the US Child Welfare system, Saxena et al. \cite{saxena_human-centered_2020} found that the literature focused mainly on risk assessment models but does not consider theoretical approaches or stakeholder perspectives. Kim et al. \cite{kim_human-centered_2021} found that incorporating human-centeredness in algorithm design can help develop more practical bullying detection systems that are better designed for the diverse needs and contexts of the stakeholders. The literature \cite{delgado_uncommon_2022} also demonstrates how  participatory approaches enabled computer scientists and lawyers to co-design Legal AI, and help align computational research and real-world, high-stakes litigation practice. Finally, Razi et al. \cite{razi_human-centered_2021} determined that a human-centered approach was necessary for identifying best practices and potential gaps, as well as setting strategic goals in the area of online sexual risk detection. 

\section{Method}
In this section, we describe our scoping criteria and processes for conducting the systematic literature search, coding, and data analysis.

\subsection{Literature Corpus}
    
The following keywords were used as search terms to identify relevant papers: '"higher education" + algorithm', '"learning analytics"', '"higher education" + AI', '"higher education" + "predictive analytics"', and '"higher education" + "machine learning"'. All searches were conducted in the ACM Digital Library between July 1-26, 2022. Our initial search returned 4,418 unique papers. Next, each paper was reviewed for inclusion/exclusion using the following criteria:
\begin{itemize}
\item The paper is peer-reviewed published work.
\item The paper contains an algorithmic approach and a technical discussion of the computational methods, predictors and outcomes employed.
\item The paper uses student data (excluding papers that only used student social media posts) limited to higher education, including university and colleges at any level, and MOOCs (massive open online courses) produced by universities or colleges and aimed at higher ed students at any level.
\item The paper was published between 2010 and 2022.
\end{itemize}

We then cross-referenced the citations of each paper to identify any additional literature that also met our inclusion criteria. We identified sixty-two relevant articles that met our inclusion criteria.

\subsubsection{Descriptive Characteristics of the Data Set}
\begin{figure}[t]
\centering
\caption{Number of Papers by Year}\label{tab:Number of Papers by Year}
\includegraphics[width=8cm]{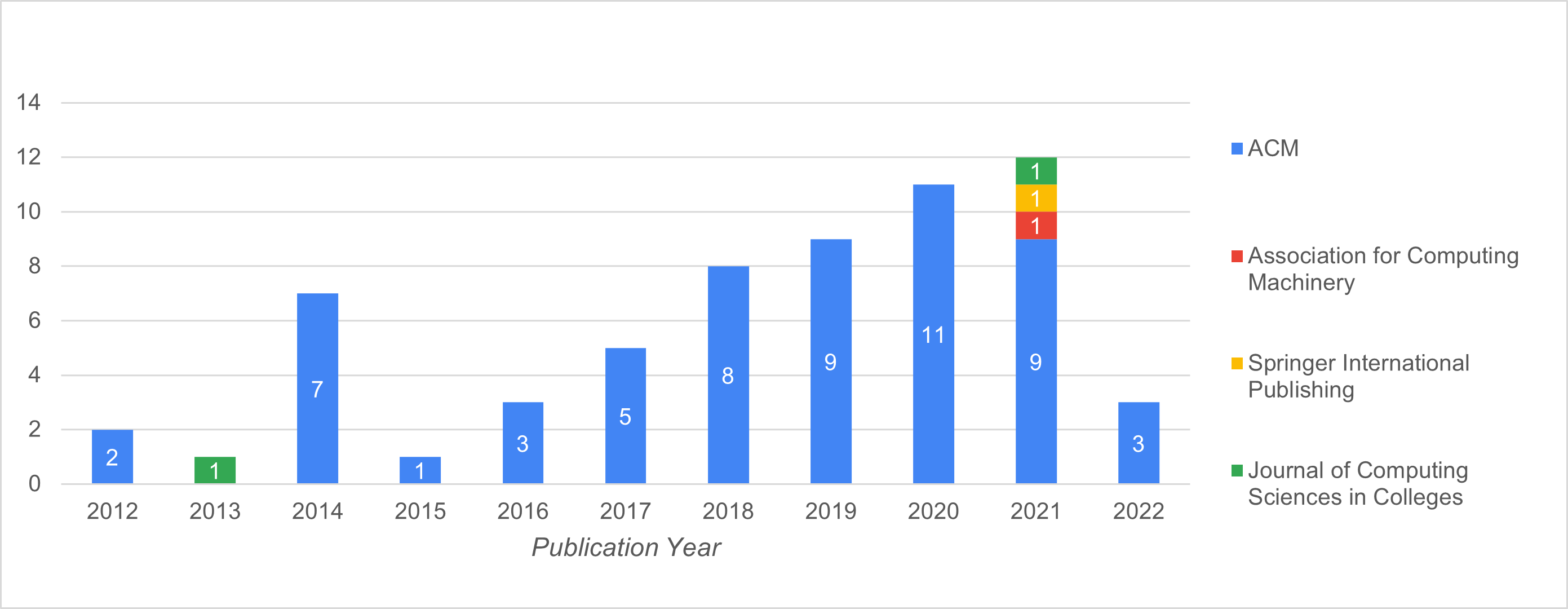}
\end{figure}

%Timeline
All of the papers reviewed (n=62) were published in the ACM full-text collection between January 2012 and July 2022 (Figure \ref{tab:Number of Papers by Year}). The majority (n=59) were published after 2013, with a significant upward trend from 2018 to 2021. We found only one paper meeting the search criteria published before 2012, a scheduling algorithm proposed by Winters in 1971 \cite{winters_scheduling_1971}. It was not included in our review. This is not to say that algorithms were not used in higher education prior to 2012. It is possible that institutions and software providers were developing and using algorithms before this time, or that research was simply published in other venues.
%Publications
The vast majority (n=59) of papers in our corpus were published as conference papers; the remainder were published in journals: Journal of Computing Sciences in Colleges \cite{pokorny_machine_2021,buerck_predicting_2013}, and Proceedings of the ACM on Interactive, Mobile, Wearable and Ubiquitous Technologies \cite{wang_first-gen_2022}. The papers came from thirty-five conferences, including the ACM Conference on Learning @ Scale (n=10) and the International Conference on Learning Analytics \& Knowledge (n=17).
\subsection{Coding and Data Analysis}
\subsubsection{Corpus Coding}
After identifying the corpus of sixty-two papers, the first author reviewed each paper for its input data sources, target outcome, and computational methods. Our analysis included close reads of the abstract, methodology, and discussion sections for each paper. Papers were reviewed for their specific data input types, target outcome variables, and statistical approaches or machine learning models; coding categories were developed from the results (Table \ref{tab:Coding Categories}). Many papers included multiple models and were therefore included in more than one category of computational method. Similarly, papers also used multiple input data sources and were categorized accordingly. Only one target outcome, however, was identified for each paper. We then conducted a quantitative analysis, determining the number of papers in each category. 

\textit{Data input type} varied greatly between papers. The features include grades, gender, race, first-generation status, prior academic achievement, and enrollment information. This list is non-exhaustive as there was no established standard or pattern for describing input data. In the case of LMS data, for example, it was described by papers as generally as "student interaction
data generated in the course" \cite{jayaprakash_open_2014} or as specifically as "Time elapsed since last click, Time spent in the course during 7 days, Clicks in time frames, Clicks to date, Clicks in the course during 7 days, Clicks in the forums, etc." \cite{borrella_predict_2019}. Institutional data includes data specific to the institution (including program or faculty-level data) as opposed to the student: e.g., enrollment and retention rates, geographic data, and financial information such as endowments and tuition rates. \textit{Target outcomes} were coded as one of the following categories: Student Services, Engagement, Admissions, Grade Prediction, Retention, Pathway Advising, Assessment, or Institutional Planning. \textit{Statistical approaches} include logistic regression, causal inference, index method, linear multiple regression, Cox Proportional Hazard Regression, Chi-Square test of association, MANOVA, and ANOVA, as well as general references to non-specific statistical techniques. \textit{Machine learning models} include Arima, Linear Regression, Decision Trees, Logistic Regression, Naïve Bayes, Forward Stepwise Regression, LASSO, Random Forest, XGBoost, Latent Dirichlet Allocation, K-Nearest Neighbors, and various neural nets, amongst others.

In order to analyze the papers in the context of Human-Centered Algorithm Design, we also deductively coded each paper as demonstrating the dimensions of theoretical design, participatory design, and/or speculative design. We adopted Baumer's human-centered algorithmic design framework \cite{baumer_toward_2017} using the following assessments:
\emph{Theoretical Design}:  
(i) How has the design of algorithms proposed by the papers in our corpus aligned with, or were led by, educational theory?
\emph{Participatory Design}: 
(i) How was meaningful inclusion of stakeholders (students, graduates, applicants, faculty, counselors, and/or administration) realized in the data selection, algorithm design, model evaluation, or implementation processes? 
(ii) How was the model evaluated using stakeholder feedback? 
\emph{Speculative Design}:
(i) How have researchers envisioned their proposed algorithms being used in real-world higher-education institutions and scenarios, including the consideration of potential harms and consequences?
\subsubsection{Data Analysis}
To determine trends, we examined the change in the size of each category over time, from 2012 to 2022. To better understand the relationships between the categories, we also cross-tabulated the papers (computational method $\leftrightarrow$ target outcome, target outcome $\leftrightarrow$ input data, and computational method $\leftrightarrow$ input data). Finally, we cross-tabulated our categories with each of the HCAD dimensions: theoretical design, participatory design, and speculative design (Table \ref{tab:Human Centered Algorithm Design Strategy by Feature, Target Outcome and Computational Method}).

\section{Results}
In this section, we %analyse 
present the results of the quantitative analysis of the paper's dataset, structured around 
%the papers in terms of the computational methods used, 
the goals of the papers, the predictive features employed, and the computational methods used. We then review the papers for dimensions of theoretical design, participatory design, and speculative design. Finally, we explore the relationships between these elements.
        \begin{table*}[]
        \centering
        \caption{Demographic Features by Year}\label{tab:Demographic Features by Year}
        %\resizebox{\columnwidth}{!}{%
        \begin{tabular}{@{}cccccccccccc@{}}
        \toprule
        \textbf{Year} & \textbf{\begin{tabular}[c]{@{}c@{}}Total\\ Papers\end{tabular}} & \textbf{Gender} & \textbf{Age} & \textbf{\begin{tabular}[c]{@{}c@{}}Prior\\ Education\end{tabular}} & \textbf{Ethnicity} & \textbf{Income} & \textbf{Location} & \textbf{\begin{tabular}[c]{@{}c@{}}Marital\\ Status\end{tabular}} & \textbf{First Gen} & \textbf{\begin{tabular}[c]{@{}c@{}}Disability\\ Status\end{tabular}} & \textbf{\begin{tabular}[c]{@{}c@{}}Personal\\ Social Media\end{tabular}} \\ \midrule
        \textit{2012} & 6 & 1 & 1 & 1 & 1 & 1 & - & 1 & - & - & - \\
        \textit{2014} & 6 & 2 & 1 & 1 & 1 & - & - & - & 1 & - & - \\
        \textit{2015} & 3 & 1 & 1 & - & - & - & - & 1 & - & - & - \\
        \textit{2016} & 6 & 1 & 1 & - & 1 & 1 & - & 1 & 1 & - & - \\
        \textit{2017} & 15 & 4 & 3 & 3 & - & 1 & 2 & 1 & - & 1 & - \\
        \textit{2018} & 5 & 2 & 2 & - & - & - & - & - & - & - & 1 \\
        \textit{2019} & 12 & 3 & 2 & 3 & 1 & - & 1 & 1 &  & 1 & - \\
        \textit{2020} & 20 & 5 & 4 & 3 & 3 & 1 & 2 & 1 & 1 & - & - \\
        \textit{2021} & 15 & 4 & 3 & 2 & 2 & 2 & 1 & - & - & - & 1 \\
        \textit{2022} & 2 & - & - & - & 1 & - & - & - & 1 & - & - \\ \midrule
        %\% of Corpus & 44\% & 37\% & 29\% & 21\% & 16\% & 10\% & 10\% & 10\% & 6\% & 3\% & 3\%
        \end{tabular}%
        %}
        \end{table*}
\subsection{Input Data}

Here we examine the predictors used to develop algorithms in higher education across six dimensions: demographic, learning management system activity (LMS), institutional, grade/GPA, enrollment/pathways, and student surveys. Three papers used for assessment were not categorized as they used assignment-specific input data \cite{sallaberry_comparison_2021, luzardo_estimation_2014, echeverria_presentation_2014}.

%Results
Twenty-eight of the sixty-two papers reviewed include student demographic data of some kind in the feature set. The specific input data are shown in Table \ref{tab:Demographic Features by Year}. The most commonly used demographic features were gender, age, prior education, and ethnicity. Two papers \cite{benablo_higher_2018,oreshin_implementing_2020} use students' personal social media activity (self-reported use of social networking sites and engagement data collected from students' personal pages) for the prediction of grades and retention, respectively. Use of household financial data is increasing, papers included as input features: family assets \cite{daud_predicting_2017}, need for financial assistance \cite{yu_should_2021}, and parental income \cite{ameri_survival_2016, barber_course_2012, jiang_towards_2021, kung_interpretable_2020}.

Five papers have models built only with unique features that could not be categorized. Three are for the purpose of assignment evaluation and use variables specific to the assignment. This includes slide specifications and audio for measuring presentation skills \cite{luzardo_estimation_2014, echeverria_presentation_2014}, and velocity, acceleration, jerk, and rotation speed in a dental simulator \cite{sallaberry_comparison_2021}. The remaining two use student behavior (absenteeism, tardiness, uniform violation, and misconduct) \cite{cabrera_data_2020} and student emails \cite{hien_intelligent_2018} as input data for student services algorithms that provide students with guidance.

%Protected Classes
In our review, we found almost half (n=27) of the papers include one or more protected attributes (defined as race, sex, age, disability status, or citizenship). Protected attributes, attributes for which discrimination is illegal or protected by some policy or authority, are commonly used as algorithmic inputs in higher education \cite{yu_should_2021}. Some of these attributes, minority status, and family income, for example, are substantially correlated with higher education dropout rates \cite{de_brey_status_2019}. However, Yu et al. \cite{yu_should_2021} found that the inclusion of protected attributes did not improve their model's performance for dropout prediction, and the use of these attributes as input data may amplify inequities already existing in higher education institutions \cite{yu_towards_2020}.

Thirty-two papers include student grades as a predictor. This data encompasses course grades, assignment grades, GPA, transcripts, and standardized test results. Student enrollment is also used frequently (n=18) in the form of input data related to course load, course descriptions, educational pathways, and currently enrolled courses.

Authors frequently use interactions with the institution's learning management system as measures of student learning behaviors and student engagement in their respective courses. Eighteen of the papers reviewed used LMS data as input data for their algorithms. Examples of LMS data include timestamps and counts of clicks, time spent watching videos or reading course content, use of discussion boards, and login frequency.

Only six of the papers reviewed include student survey data. The surveys include perceptions of the LMS \cite{maramag_assessing_2019}, self-reported mental health data \cite{wang_first-gen_2022}, skills, interests, and preferences \cite{obeid_ontology-based_2018}, student services needs \cite{hien_intelligent_2018}, self-reported learning styles \cite{mcpherson_student_2016} and prior knowledge assessment \cite{edwards_using_2017}. Finally, few of the reviewed papers (n=3) had input data related to the institution or course more generally, and all three were published in 2021. These features included enrollment and retention \cite{pokorny_machine_2021, zhou_analysis_2021, amballoor_technological_2021}, and digital competency of the institution and faculty \cite{amballoor_technological_2021}. 

%Importance
The feature selection process is an important element of model design. Guyon et al. \cite{guyon_introduction_2003} identify three main objectives of variable selection: improved model performance, improved model efficiency (both speed and cost in acquiring and cleaning data, and in running the model), and model explainability. The papers in our review overwhelmingly prioritized model performance and data availability. Explainability was not detailed as part of the feature selection process nor was the process grounded in educational theory. A human-centered approach to algorithm design should begin with a contextual understanding of the data \cite{cherrington_features_2020} as the curation of input data creates opportunities for bias, algorithmic harm, and privacy concerns. The features selected become the 'ground truth' of the model but the feature selection process is inherently subject to interpretation \cite{cherrington_features_2020}. In the majority of the papers in our review, the roles of gender, age, and ethnicity became the 'ground truth' in predicting grades or student success. The apparent correlation between these personal attributes and a student's academic achievements is unquestioned, and few papers adopted a human-centered lens to the selection process. Missing from the papers is evidence of an evaluation of the features through educational or social science theory, to speculate on the possible impact of using these features in decision-making, or to include stakeholders in the feature selection process.
        
    \subsection{Computational Methods}
In this section, we discuss the computational methods used to develop algorithms. For the purpose of analysis, methods and model types were categorized as Inferential Statistics, Rules Based, Machine Learning (ML), Deep Learning (DL), and Natural Language Processing (NLP).

%Model Selection Findings
The use of machine learning algorithms within higher education grew rapidly beginning in 2014, as shown in Table \ref{tab:Computational Method by Year}. As expected, inferential statistics were primarily used prior to this trend. As machine learning methods advanced, researchers moved away from statistical methods. 

The overwhelming majority of papers in our corpus, 48 (77\%), used some form of machine learning. Almost half of the papers (n=29) used more than one type of machine learning algorithm (46\% of all papers, and 60\% of machine learning papers). These papers have a stated goal of comparing models for the task at hand.

From 2018 to present, NLP and deep learning both saw rapidly increasing use, indicating a shift towards less explainable and less interpretable models and results. Only four of these papers included any discussion of interpretability and explainability, including feature importance. To increase the interpretability of recurrent neural networks and determine feature importance, authors used the permutation feature importance algorithm  \cite{wang_first-gen_2022}, random forests \cite{altaf_student_2019} and SHAP global feature importance \cite{baranyi_interpretable_2020}. While these methods offer insight into how the algorithm's decision was weighted, only one \cite{baranyi_interpretable_2020} discussed potential interpretations of the results.
        
        \begin{table}[]
        \centering
        \caption{Computational Method by Year}\label{tab:Computational Method by Year}
        \resizebox{\columnwidth}{!}{%
        \begin{tabular}{ccccccc}
        \hline
        \textbf{Year} & \textbf{\begin{tabular}[c]{@{}c@{}}Total \\ Papers\end{tabular}} & \textbf{\begin{tabular}[c]{@{}c@{}}Inferential \\ Statistics\end{tabular}} & \textbf{\begin{tabular}[c]{@{}c@{}}Rules\\ Based\end{tabular}} & \textbf{\begin{tabular}[c]{@{}c@{}}Machine\\ Learning\end{tabular}} & \textbf{\begin{tabular}[c]{@{}c@{}}Deep\\ Learning\end{tabular}} & \textbf{\begin{tabular}[c]{@{}c@{}}Natural\\ Language\\ Processing\end{tabular}} \\ \hline
        %\textit{1971} & 1 & 1 & - & - & - & - \\
        \textit{2012} & 2 & 1 & - & 1 & - & - \\
        \textit{2013} & 1 & 1 & - & - & - & - \\
        \textit{2014} & 7 & 3 & - & 5 & - & - \\
        \textit{2015} & 1 & - & - & 1 & - & - \\
        \textit{2016} & 3 & 2 & 1 & - & - & - \\
        \textit{2017} & 5 & 2 & - & 3 & - & - \\
        \textit{2018} & 8 & - & 1 & 6 & 2 & 1 \\
        \textit{2019} & 9 & 2 & - & 5 & 4 & 1 \\
        \textit{2020} & 11 & 1 & - & 9 & 3 & 1 \\
        \textit{2021} & 11 & 1 & - & 8 & 3 & 1 \\
        \textit{2022} & 4 & - & - & 3 & 2 & 1 \\ \hline
        \end{tabular}
        }
        \end{table}

%Model Selection Impacts HCAD
Computational methods can have a meaningful impact on the potential for algorithmic harm. In the case of machine learning, model selection affects more than just the algorithmic performance; explainability and interpretability are key factors ensuring algorithmic decision-making is as fair as possible and human-centered \cite{kung_interpretable_2020}. As Rudin \cite{rudin_stop_2019} noted, there remains a pervasive myth in the research community that the model selection process inevitably includes a trade-off between model accuracy and interpretability, "a widespread belief that more complex models are more accurate, meaning that a complicated black box is necessary for top predictive performance." We saw this belief shape research in many of the papers we reviewed; with model performance as the primary metric and goal, explainability and interpretability are dismissed in favor of deep learning methods. In order to develop human-centered algorithms for higher education, students must be centered in the model selection process by prioritizing results that can be understood and explained.

\subsection{Target Outcomes}
In this section, we examine the target outcomes of algorithms proposed in the papers. The papers were sorted into nine categories based on the goal of the model and the target variable: grade prediction, retention, institutional planning, pathway advising, student services, admissions, assessment, and engagement. Table \ref{tab:Target Outcome by Year} depicts the distribution of papers across target outcome categories, by year.
 %\cite{hien_intelligent_2018,rijati_multi-attribute_2018,wang_first-gen_2022,fischer_mining_2020}

Thirty-nine (63\%) of the papers in our review propose algorithms that seek to predict student success, defined as retention or dropout prediction (n=19), grade prediction (n=18), and admissions decision-making (n=2). While still a significant focus of the research community, student success papers have accounted for a smaller share of the research in recent years. 

Student services (n=4) and pathway advising (n=7) make up 18\% of the papers. The increase in available data from learning management systems and self-serve course registration may account for the recent increase in research in these two areas.

      \begin{table*}[]
        \centering
        \caption{Target Outcome by Year}\label{tab:Target Outcome by Year}
        %\resizebox{\columnwidth}{!}{%
        \begin{tabular}{@{}ccccccccccc@{}}
        \toprule
        \textbf{Year} & \textbf{\begin{tabular}[c]{@{}c@{}}Total\\ Papers\end{tabular}} & \textbf{\begin{tabular}[c]{@{}c@{}}Grade\\ Prediction\end{tabular}} & \textbf{Retention} & \textbf{\begin{tabular}[c]{@{}c@{}}Institutional\\ Planning\end{tabular}} & \textbf{\begin{tabular}[c]{@{}c@{}}Pathway\\ Advising\end{tabular}} & \textbf{\begin{tabular}[c]{@{}c@{}}Student\\ Services\end{tabular}} & \textbf{Admissions} & \textbf{Assessment} & \textbf{Engagement} \\ \midrule
        %\textit{1971} & 1 & - & - & 1 & - & - & - & - & - \\
        \textit{2012} & 2 & 1 & 1 & - & - & - & - & - & - \\
        \textit{2013} & 1 & 1 & - & - & - & - & - & - & - \\
        \textit{2014} & 7 & 3 & 2 & - & - & - & - & 2 & - \\
        \textit{2015} & 1 & - & 1 & - & - & - & - & - & - \\
        \textit{2016} & 3 & 1 & 1 & - & - & - & - & - & 1 \\
        \textit{2017} & 5 & 2 & 3 & - & - & - & - & - & - \\
        \textit{2018} & 8 & 2 & 1 & 1 & 1 & 2 & - & - & 1 \\
        \textit{2019} & 9 & 2 & 3 & - & 3 & - & - & - & 1 \\
        \textit{2020} & 11 & 3 & 4 & - & 1 & 1 & 2 & - & - \\
        \textit{2021} & 11 & 3 & 2 & 4 & 1 & - & - & - & 1 \\
        \textit{2022} & 4 & - & 1 & - & 1 & 1 & - & 1 & - \\ \hline
        \end{tabular}%
        %}
        \end{table*}

The shift in focus from student success prediction for retention towards pathways advising and admissions is significant. Retention and drop-out prediction models are proposed to provide insight to academic staff by identifying students in need who may not have been otherwise supported. Pathway advising and, to an even greater degree, admissions models have the power to make decisions for students, acting as gatekeepers to courses and programs. 

Seven papers focused on course-level predictions, including predicting student engagement (n=4) and student assessment (such as automated assignment evaluation) (n=3). This is another growing area of research that is also dependent on the influx of LMS data. The remaining five papers related to institutional planning. There is growing variability in the goals of the models proposed by the research community. Ten years ago, models were largely limited to grade prediction and retention. In the last few years, there has been a trend towards student service models: models that students interact with directly for information and advice, such as pathway advising, administration services, and learning support.

\subsection{Human-Centered Algorithm Design Strategies}

The vast majority of the literature in our review does not include educational, social, or behavioral science theories in the design of their algorithms. Only two of the papers ground their algorithmic design in established learning theory: Borella et al. \cite{borrella_predict_2019} consider individualistic and constructivist framings while designing interventions for their drop-out prevention tool, and interest exploration, which the authors consider a fundamental component of constructivist learning, is critical to the design of the educational pathways tool developed by Chen et al. \cite{chen_pathways_2022}. All sixty-two papers in our corpus focus on students and learning, but despite the subject matter, only these two papers make substantial reference to educational or learning theory. 

Despite the participatory nature of higher education, only three papers in our review  \cite{fiorini_application_2018,pardos_designing_2020,chen_pathways_2022} include a participatory approach to their model development. All three have pathway advising as their targeted outcome. The participatory strategies employed include a formative study on need analysis \cite{chen_pathways_2022}, post-intervention evaluation studies \cite{chen_pathways_2022, pardos_designing_2020}, and a robust participatory action research approach directly involving academic advisors \cite{fiorini_application_2018}.

While no papers demonstrate a robust speculative design, three papers \cite{kung_interpretable_2020,yu_should_2021,jiang_towards_2021} do consider the use of the tool in serving real-world purposes. For all three, that consideration includes warnings of potential consequences and the "broader implications for using predictive analytics in higher education" \cite{yu_should_2021}. One paper suggests real-world mitigation techniques by advocating for human-in-the-loop implementation \cite{kung_interpretable_2020}.
    
\subsection{Relationship between Input Data, Methods, Outcomes, and HCAD Strategies}
In this section, we examine the trends in the interactions between the input data, computational methods, and target outcomes. We then discuss the relationship between each of the model parameters and each of the human-centered algorithm design strategies: theoretical design, participatory design, and speculative design.\\
\textbf{Relationship between Input Data and Target Outcomes}

Table \ref{tab:Input Data and Computational Methods by Target Outcome} crosstabs between the input data used and the target outcomes they seek to predict. GPA/Grades are used as a predictor for all target outcomes at least once, with the sole exception of algorithms developed to assess individual assignments. Demographic data, which includes the protected attributes, is the next most commonly employed predictor and weighs heavily for grade prediction and retention. LMS activity is applied not only in predicting engagement but also for grade prediction, pathway advising, and retention. Two papers use student surveys for the purpose of grade prediction, a self-report Learning Style Inventory survey \cite{bos_student_2016} and a prior knowledge survey \cite{edwards_using_2017}.

\noindent\textbf{Relationship between Computational Method and Target Outcome}

The computational method used and the correlating target outcome are also shown in Table \ref{tab:Input Data and Computational Methods by Target Outcome}. The review revealed a high variety of computational methods across target outcomes, except in the case of admission prediction (machine learning only). Within those two papers \cite{alghamdi_machine_2020, staudaher_predicting_2020} however, a variety of machine learning algorithms are employed. Papers with the most common target outcomes, Grade Prediction, and Retention, include statistical, machine learning, and deep learning methods. Natural Language Processing, which is only found in papers published in 2018 or later, is used only for Institutional Planning, Pathway Advising, and Student Services.

\noindent\textbf{Relationship between Input Data and Computational Method}

Next, we cross-examine input data by computational method (Table \ref{tab:Input Data by Computational Method}). Papers employing statistical methods and machine learning techniques both used a wide variety of features as predictors. This is to be expected of machine learning, a modeling technique that requires heavily engineered feature sets and, within the papers reviewed here, is used to predict a very wide variety of target outcomes. Statistical methods, however, are less expected as the papers using this technique are limited to four target outcomes. For Grade Prediction, papers employing statistical methods use input data across five categories. For the same target outcome, papers employing deep learning methods use only demographic, GPA, and, for a singular paper, enrollment data. The most recent technique, Natural Language Processing, is associated with only three feature categories: GPA (1 paper \cite{chockkalingam_which_2021}), student survey (1 paper \cite{hien_intelligent_2018}), and enrollment/pathway data (4 papers all using course descriptions \cite{chockkalingam_which_2021,pardos_designing_2020, chen_pathways_2022, pardos_data-assistive_2019}). 

\noindent\textbf{Relationship between HCAD Strategy and Predictor, Target Outcome and Computational Method}

In addition to the model parameters described previously, our review also considered the design of the study through the lens of Baumer's \cite{baumer_toward_2017} framework described in section \ref{ssec:Human-Centered Algorithm Design}. In this section, we'll identify which papers employ theoretical, participatory, or speculative dimensions in their design, as demonstrated in Table \ref{tab:Human Centered Algorithm Design Strategy by Feature, Target Outcome and Computational Method}. In the next section, \ref{sec:Discussion}, we'll discuss the impacts of this inclusion on the literature.

Only one paper \cite{chen_pathways_2022} included more than one element of Baumer's HCAD framework \cite{baumer_toward_2017} (theoretical and participatory design). No papers to our knowledge include all three dimensions. All feature categories except \textit{Institutional} and \textit{Student Survey} were accounted for in papers using HCAD strategies. Interestingly, while three papers \cite{fiorini_application_2018,pardos_designing_2020,chen_pathways_2022} using participatory design included participant surveys in their design methodology, they did not include surveys as feature sets. Only one paper \cite{hien_intelligent_2018} used student survey data in its feature set, in the design of a student services chatbot. The HCAD papers generally targeted the more common outcomes of \textit{Grade Prediction} and \textit{Retention}, though two papers \cite{pardos_designing_2020,chen_pathways_2022} were focused on \textit{Pathway Advising}. All of the HCAD papers employed machine learning techniques including machine learning (4 papers), deep learning (2 papers), and NLP (3 papers).
\section{Discussion} \label{sec:Discussion}
%In this section, we discuss the trends and gaps identified in our corpus. We also discuss the challenges with these trends, the use of human-centered lenses within the literature, and opportunities for future research.

\subsection{Research Challenges and Opportunities for Algorithmic Decision-Making in Higher Education}
Three major needs emerged from the trends identified by our review: the need to establish valid and theory-informed ground truths, the need to consider interventions within the algorithm design process, and the need for governance as algorithms move from identifying and informing to decision-making.

\begin{table*}[]
\centering
\caption{Input Data and Computational Method by Target Outcome}\label{tab:Input Data and Computational Methods by Target Outcome}
%\resizebox{\columnwidth}{!}{%
\begin{tabular}{lccccccccccc}
 & \multicolumn{6}{c}{\textbf{Input Data}} & \multicolumn{5}{c}{\textbf{Computational Method}} \\ \hline
\multicolumn{1}{c|}{\textbf{Target Outcomes}} & \textit{DM} & \textit{LMS} & \textit{IN} & \textit{GPA} & \textit{ENR} & \multicolumn{1}{c|}{\textit{SS}} & \textit{SM} & \textit{RB} & \textit{ML} & \textit{DL} & \textit{NLP} \\ \hline
\multicolumn{1}{l|}{\textit{Admissions}} & 1 & - & - & 1 & - & \multicolumn{1}{c|}{-} & - & - & 2 & - & - \\
\multicolumn{1}{l|}{\textit{Assessment}} & - & - & - & - & - & \multicolumn{1}{c|}{-} & - & - & 3 & 1 & - \\
\multicolumn{1}{l|}{\textit{Engagement}} & 1 & 3 & - & 2 & 1 & \multicolumn{1}{c|}{-} & - & 1 & 3 & - & - \\
\multicolumn{1}{l|}{\textit{Grade Prediction}} & 10 & 8 & - & 13 & 4 & \multicolumn{1}{c|}{2} & 6 & - & 11 & 6 & - \\
\multicolumn{1}{l|}{\textit{Institutional Planning}} & - & - & 3 & 2 & 3 & \multicolumn{1}{c|}{-} & 1 & 1 & 3 & 1 & 1 \\
\multicolumn{1}{l|}{\textit{Pathway Advising}} & - & 1 & - & 2 & 5 & \multicolumn{1}{c|}{1} & 1 & - & 3 & 2 & 3 \\
\multicolumn{1}{l|}{\textit{Retention}} & 15 & 6 & - & 11 & 5 & \multicolumn{1}{c|}{1} & 4 & - & 14 & 2 & - \\
\multicolumn{1}{l|}{\textit{Student Services}} & 1 & - & - & 1 & - & \multicolumn{1}{c|}{2} & - & - & 2 & 2 & 1 \\ \hline
 & \multicolumn{6}{l}{\textbf{DM:} Demographics} & \multicolumn{5}{l}{\textbf{SM:} Statistical Methods} \\
 & \multicolumn{6}{l}{\textbf{LMS:} LMS/Engagement} & \multicolumn{5}{l}{\textbf{RB:} Rules-Based} \\
 & \multicolumn{6}{l}{\textbf{IN:} Institutional} & \multicolumn{5}{l}{\textbf{ML:} Machine Learning} \\
 & \multicolumn{6}{l}{\textbf{GPA:} Grade/GPA} & \multicolumn{5}{l}{\textbf{DL:} Deep Learning} \\
 & \multicolumn{6}{l}{\textbf{ENR:} Enrollment/Pathways} & \multicolumn{5}{l}{\textbf{NLP:} Natural Language} \\
 & \multicolumn{6}{l}{\textbf{SS:} Student Survey} & \multicolumn{5}{l}{Processing} 
\end{tabular}
\end{table*}
\subsubsection{Establishing Theoretical Groundedness}
We found a growing trend toward the use of Learning Management System (LMS) data as input features in algorithm design (n=18). These models were built with the goals of predicting grades (n=8), retention (n=6), engagement (n=3), and pathway advising (n=1). In each of these papers, student interaction with their web-based learning platform was used as a proxy for their engagement in the course. The underlying assumption is that LMS activity like clicks on videos, frequency of logins, and views of course content is correlated with their interest and participation in the material, yet in none of the papers was this assumption questioned or was educational theory presented to support it. LMS interaction as a proxy for engagement, and an indicator of academic success, is then accepted as a ground truth within the algorithms. LMS activity, though, can take many forms and there was no discussion within the papers as to the predictive power of specific types of activities. The viewing of a video may not have equal engagement value as participation in a forum discussion, or infrequent but substantial logins may or may not be as effective as frequent, brief ones. Additionally, activity can be impacted by many factors, including technological factors such as interface design and privacy support \cite{zanjani_important_2017} and the combination of the LMS itself, the course design, and the course type produce different student experiences and influence student behaviors \cite{demmans_epp_learning_2020}. These factors could differ significantly from course to course, even within the same program or institution. The algorithms in question don't account for the course design. The ground truth of these models is that students with less LMS activity are less likely to succeed, the output of which will trigger a student-level intervention and not a review of the LMS, course design, or user experience.

Beyond incorporating educational theory in the design of algorithms, computational scientists should allow for more collaboration with education subject matter experts and social scientists, to analyze the importance of these LMS interactions critically. More research is needed to understand the relationship between LMS activity and engagement. Additionally, researchers must consider how the importance of the individual features of their algorithms aligns not just with target variables, but also with the greater goals and outcome of the algorithm. We discuss the importance of proposing appropriate interventions as part of the algorithm design process further in the next section.
\subsubsection{Considering Interventions within Algorithm Design}
We found that before 2018, research focused primarily on designing algorithms to predict students' grades and retention. The motivation behind this research is to reduce attrition through the early detection of students who are identified as likely to drop out or fail a course or program. Once identified, higher education institutions can provide these students with timely and focused interventions. Papers frequently cite significant consequences to students, academic staff, and higher education institutions as a result of attrition, but few consider what interventions exist, the efficacy of the interventions, or the institutions' legal and ethical obligations to provide interventions once 'at-risk' students have been identified \cite{prinsloo_elephant_2017}. Intervention efficacy, as well, is difficult to measure - interventions varied in form and included automated email notifications, supplementary readings and assignments, recurring meetings with advisors, and counseling services \cite{harackiewicz_improving_2018}. The one paper identified that did consider the impact of these interventions \cite{dawson_prediction_2017} found no evidence of a subsequent effect on retention outcomes. While dropout prediction was and remains a popular avenue of research, we did not find any literature demonstrating that the models are put into use effectively within higher education institutions. The algorithms seem to be designed within data-science silos, with little consideration for their role within the institution or as part of the greater student experience. A participatory design approach that incorporates learning strategists, for example, allows for the iterative design of theoretically grounded interventions in tandem with the development of the algorithm; any limitations and potentially harmful impact of the model (such as false positives) can be explained and considered in the design of the interventions \cite{brooks_explaining_2014}.

While more research is certainly needed to determine if algorithms developed to identify 'at-risk' students and subsequently trigger early interventions have any impact on retention, the acute risk to students appears minimal: at worst the resulting outcome is participation in an ineffective intervention. Student support services at higher education institutions can be scarce, however. Support that goes to a student flagged algorithmically may be support that would have otherwise been provided to another student. In other words, within higher education institutions, interventions may be a finite resource. We were unable to find any literature that compared the outcomes for students identified by an algorithm to those identified through traditional methods.

To address these gaps, future research must include consideration of the interventions proposed as an outcome of the model. An algorithm designed to increase retention cannot only be evaluated by its performance in predicting which students may not succeed. In other words, researchers should consider the relationship between the specific predictors used in making the prediction and the intervention proposed to then correct the prediction.

\subsubsection{Moving from Identification to Decision-Making}
Even more concerning, however, is the move from retention and grade prediction toward pathway advising and admissions. The decisions made by retention algorithms create access to support services that may not have otherwise been available to students but they do not restrict students' choices. Algorithms designed for course selection, program admission, and pathway advising have the potential for a more direct and limiting impact on students. For example, researchers \cite{staudaher_predicting_2020} used the same algorithmic target variable, retention, along with likelihood of job placement, to develop an admissions algorithm. In this case, retention was not used to flag enrolled students in need of assistance, but rather to deny applicants admission to a Master's program. In another example, researchers \cite{jiang_time_2019} used grade history to build a course recommendation engine, guiding students in what courses to pursue based on likelihood of success. In both these cases, grade prediction is used as a tool to restrict students' options. Algorithms in higher education are moving beyond the ability to simply provide greater insight to institutions through the identification of at-risk students, but to actually make the decisions for them.

As we expect to see this trend continue, and the use of algorithms for decision-making expand beyond its current scope, our results suggest the need for more governance within higher education institutions. Our review indicates that it is certainly technologically possible to leave these student decisions in the hands of algorithms, but whether it is desirable for algorithms to shape students' academic careers without the oversight of a human-in-the-loop remains a legal, ethical, and pedagogical question.
\begin{table}
\centering
\caption{Input Data by Computational Method}\label{tab:Input Data by Computational Method}
\begin{tabular}{lcccccc}
\multicolumn{1}{l}{}                                     & \multicolumn{6}{c}{\textbf{Input Data}}                                                 \\ \hline
\multicolumn{1}{l|}{\textbf{Computational Method}}        & \textit{DM} & \textit{LMS} & \textit{IN} & \textit{GPA} & \textit{ENR} & \textit{SS} \\ \hline
\multicolumn{1}{l|}{\textit{Statistical Methods}}         & 5           & 7            & 1           & 6            & 2            & 3           \\
\multicolumn{1}{l|}{\textit{Rules-Based}}                 & -           & 1            & -           & 1            & -            & -           \\
\multicolumn{1}{l|}{\textit{Machine Learning}}            & 21          & 9            & 2           & 21           & 12           & 2           \\
\multicolumn{1}{l|}{\textit{Deep Learning}}               & 6           & 2            & -           & 9            & 5            & 1           \\
\multicolumn{1}{l|}{\textit{Natural Language Processing}} & -           & -            & -           & 1            & 4            & 1           \\ \hline
                                                          & \multicolumn{6}{l}{\textbf{DM:} Demographics}                                                 \\
                                                          & \multicolumn{6}{l}{\textbf{LMS:} LMS/Engagement}                                              \\
                                                          & \multicolumn{6}{l}{\textbf{IN:} Institutional}                                                \\
                                                          & \multicolumn{6}{l}{\textbf{GPA:} Grade/GPA}                                                   \\
                                                          & \multicolumn{6}{l}{\textbf{ENR:} Enrollment/Pathways}                                         \\
                                                          & \multicolumn{6}{l}{\textbf{SS:} Student Survey}                                              
\end{tabular}
\end{table}
\subsection{Towards Human-Centeredness in Algorithmic Decision-Making in Higher Education}
Without the application of a human-centered lens, much of the discussion above would have been missed. Embedding Baumer's framework \cite{baumer_toward_2017} enabled us to center students within our review, and subsequently identify the limitations of the papers themselves and the increased potential for harm within the changing trends of algorithm design. 

Higher education institutions are socially-complex, diverse, and high-stakes environments with a potentially vulnerable population of students and inherent power imbalance. Algorithmic systems designed to provide insights into, or more recently, to make decisions for, the student population are inherently part of these systems. Complex and highly contextual student data is used to train algorithms that are then used to make complex decisions for those students. The research community has raised concerns about the use of AI that includes machine autonomy, the consequences of which are quickly being realized \cite{shneiderman_human-centered_2021}. The AI Incident Database strives to document these risks and impacts, and includes many examples of algorithmic harm in higher education including facial-recognition software for exam proctoring providing "allegedly discriminatory experiences for BIPOC students" \cite{hall_incident_2020} and wrongfully accusing students of academic dishonesty, an application screening algorithm that "allegedly exacerbated existing inequality for marginalized applicants" \cite{hall_incident_2012}, and a grading algorithm that "kept students out of college" \cite{stockton_incident_2020}.
\begin{table*}
\centering
\caption{Human Centered Algorithm Design Strategy by Target Outcome, Input Data, and Computational Method}\label{tab:Human Centered Algorithm Design Strategy by Feature, Target Outcome and Computational Method}
%\resizebox{\columnwidth}{!}{%
%\begin{tabular}{cl111}
\begin{tabular}{cllll}
\multicolumn{1}{l}{}                                                                      &                                      & \multicolumn{3}{c}{\textbf{HCAD Strategy}}                    \\ \hline
\multicolumn{1}{l}{}                                                                      &                                      & \textit{Theoretical} & \textit{Participatory} & \textit{Speculative} \\ \hline
\multirow{2}{*}{\textbf{\begin{tabular}[c]{@{}c@{}}Target \\ Outcomes\end{tabular}}}      & \textit{Admissions}                  & -                    & -                      & -                    \\
                                                                                          & \textit{Assessment}                  & -                    & -                      & -                    \\
\textbf{}                                                                                 & \textit{Engagement}                  & -                    & -                      & -                    \\
\textbf{}                                                                                 & \textit{Grade Prediction}            & -                    & -                      & 2 \cite{kung_interpretable_2020, jiang_towards_2021}                \\
\textbf{}                                                                                 & \textit{Institutional Planning}      & -                    & -                      & -                    \\
\textbf{}                                                                                 & \textit{Pathway Advising}            & 1 \cite{chen_pathways_2022}                   & 2 \cite{pardos_designing_2020,chen_pathways_2022}                  & -                    \\
\textbf{}                                                                                 & \textit{Retention}                   & 1 \cite{borrella_predict_2019}                & 1 \cite{fiorini_application_2018}                  & 1 \cite{yu_should_2021}                   \\
\textbf{}                                                                                 & \textit{Student Services}            & -                    & -                      & -                    \\ \hline
\textbf{Input Data}                                                                          & \textit{Demographic}                 & -                    & -                      & 3 \cite{kung_interpretable_2020,yu_should_2021, jiang_towards_2021}            \\
\textbf{}                                                                                 & \textit{LMS/Engagement}              & 1 \cite{borrella_predict_2019}                & -                      & 1 \cite{kung_interpretable_2020}             \\
\textbf{}                                                                                 & \textit{Institutional}               & -                    & -                      & -                    \\
\textbf{}                                                                                 & \textit{Grade/GPA}                   & 1 \cite{borrella_predict_2019}                & 1 \cite{fiorini_application_2018}                     & 3 \cite{kung_interpretable_2020,yu_should_2021, jiang_towards_2021}               \\
\textbf{}                                                                                 & \textit{Enrollment/Pathways}         & 2 \cite{borrella_predict_2019,chen_pathways_2022}                    & 3 \cite{fiorini_application_2018,pardos_designing_2020,chen_pathways_2022}                     & 2 \cite{yu_should_2021, jiang_towards_2021}                \\
\textbf{}                                                                                 & \textit{Student Survey}              & -                    & -                      & -                    \\ \hline
\multirow{2}{*}{\textbf{\begin{tabular}[c]{@{}c@{}}Computational \\ Method\end{tabular}}} & \textit{Statistical Methods}         & -                    & -                      & -                    \\
                                                                                          & \textit{Rules-Based}                 & -                    & -                      & -                    \\
                                                                                          & \textit{machine Learning}            & 1 \cite{borrella_predict_2019}                & 1 \cite{fiorini_application_2018}                & 2 \cite{kung_interpretable_2020,yu_should_2021}                   \\
                                                                                          & \textit{Deep Learning}               & -                    & 1 \cite{pardos_designing_2020}                     & 1 \cite{jiang_towards_2021}                   \\
                                                                                          & \textit{Natural Language Processing} & 1 \cite{chen_pathways_2022}                   & 2 \cite{pardos_designing_2020,chen_pathways_2022}                & -                   
\end{tabular}
\end{table*}

A human-centered approach is required to ensure the algorithm is designed with an understanding of those contexts and the real-world functioning of algorithmic decisions must be grounded in social science theories. By including social science theory in the design of their algorithms, researchers can create more robust feature selection processes. Educational theory allows us to more efficiently narrow down potential features to only those that are pedagogically sound, avoid bias, and align to outcomes. Educational theories will also provide a descriptive framework, improving the interpretation of input data and outcomes. Additionally, they provide new avenues for measuring model results beyond model performance. Theoretical design is by nature multi-disciplinary, requiring the expertise of multiple and varied domains. In practical terms, this means that researchers must turn to behavioral and educational theory to understand how their stakeholders work and learn, and use that understanding to select their feature sets, interpret the results, and recommend the corresponding algorithms. 

Participatory design is frequently proposed within critiques of machine learning as a strategy for identifying and mitigating risks \cite{delgado_uncommon_2022}, within the public service\cite{saxena_conducting_2020} and other workplaces \cite{fox_worker-centered_2020}. Developing an algorithm to make decisions for students requires the formalization of that decision. As such, model developers must take care to ensure that formalization reflects the true goals of all stakeholders \cite{abebe_roles_2020}. Abebe et al. see this as an opportunity for institutional reflection, offering the opportunity for non-technical stakeholders to explicitly consider how decisions should be made \cite{abebe_roles_2020}. By including domain stakeholders, researchers can develop systems that are more fit for real-world use in socially complex contexts, such as those in higher education institutions. Moving towards a participatory design approach will allow for the active involvement of students, faculty, counselors, and administration in the model design process. Engaging these stakeholders throughout the design process will enable algorithms to be designed to integrate properly into existing systems. Including staff of higher education institutions, such as counselors and retention teams, in the design of drop-out predictions algorithm will ensure the model results and features of high importance align with the available interventions. It will also allow for a better understanding of input data, thus improving the interpretation of model results. 

The majority of papers in our review evaluated the models on their performance, usually their accuracy, and didn't consider if they delivered on their purpose. Beyond design and development, a participatory approach to model evaluation enables researchers to go beyond performance metrics, evaluating models not just on their technical accuracy but on their ability to meet the goals of the research. The algorithms may accurately identify the students it has been trained are 'at risk', but whether that identification led to meaningful interventions that subsequently increased retention is unknown. Many questions around model value, as opposed to model performance, are unanswered: do drop-out prediction models improve retention; does LMS data reliably correlate to the student experience; are students satisfied with the courses selected for them through a pathway advising tool? Without the perspectives of a participatory approach, future research will be limited to defining success by an algorithm's ability to predict that which it was trained to predict. This is, of course, a crucial step in model development. But to reduce the potential for algorithmic harm, researchers must all consider the success of their models in the context of their intended complex systems. This evaluation requires the participation of stakeholders both during the design of the model and after it is implemented.

Every algorithm in our review was proposed for some real-world purpose, from triggering interventions for at-risk students and reducing attrition to grading assignments. But the literature largely fails to consider the algorithms as part of larger systems, and how they will affect those systems going forward. The research included in our review considers model performance only, with questions of how the decisions will be governed and implemented unanswered, and little speculation about how automated decisions could shape student experiences and outcomes. Future research into algorithmic decision-making in higher education must go beyond model performance to consider the potential the algorithms could have once used for their proposed real-world purpose. To design algorithms that have the intended impact, researchers must consider both the possibility of harm caused by the model but also what changes, if any, are necessary to ensure the algorithm is able to help meet institutional goals.

\subsection{Limitations and Future Work}
Our systematic literature review was limited to papers published in the ACM Digital Library between 2010 and 2022. We may have missed algorithms published outside this time period, as well as in other libraries and those that are not publicly available as research, for example, drop-out prediction algorithms developed for profit by educational software providers. We plan to work with stakeholders in education, including students, counselors, faculty, and administration, to better understand how the algorithms in our review, and those developed proprietarily, are being implemented in higher education institutions. To move towards a human-centered approach to building evidence-based and theoretically-driven algorithms that do as they intend in real-world scenarios, we plan to seek a better understanding of the connections between input features and proposed interventions, and LMS data and student engagement. 

\section{Conclusion}

We conducted a systematic review of existing literature published in the ACM Digital Library on algorithmic decision-making in higher education. After establishing a corpus of 63 papers, we quantitatively analyzed the papers for their input data, computational methods, and target outcomes, before applying an established human-centered algorithm design framework \cite{baumer_toward_2017}. Going forward, we recommend that the HCI research community focus on developing algorithm design processes with theoretical, participatory, and speculative dimensions. Theoretically-grounded algorithm design that includes active engagement of stakeholders through the design and evaluation phases will ensure algorithms that are better aligned to the socially-complex systems of which they are a part. It will also increase understanding of highly-contextual input data and allow for better interpretation of results. Additionally, we recommend that future research in this area considers the future of higher education institutions and how the proposed algorithms may impact stakeholders.

\begin{acks}
This research was supported by the Natural Sciences and Engineering Research Council of Canada's Early Career Discovery Grant. We would like to thank the anonymous reviewers for helping us improve this work.
\end{acks}

\bibliographystyle{ACM-Reference-Format}
\bibliography{main}


\begin{thebibliography}{118}

%%% the Web address.

\ifx \showCODEN    \undefined \def \showCODEN     #1{\unskip}     \fi
\ifx \showDOI      \undefined \def \showDOI       #1{#1}\fi
\ifx \showISBNx    \undefined \def \showISBNx     #1{\unskip}     \fi
\ifx \showISBNxiii \undefined \def \showISBNxiii  #1{\unskip}     \fi
\ifx \showISSN     \undefined \def \showISSN      #1{\unskip}     \fi
\ifx \showLCCN     \undefined \def \showLCCN      #1{\unskip}     \fi
\ifx \shownote     \undefined \def \shownote      #1{#1}          \fi
\ifx \showarticletitle \undefined \def \showarticletitle #1{#1}   \fi
\ifx \showURL      \undefined \def \showURL       {\relax}        \fi
% invisible to TeX
\providecommand\bibfield[2]{#2}
\providecommand\bibinfo[2]{#2}
\providecommand\natexlab[1]{#1}
\providecommand\showeprint[2][]{arXiv:#2}

\bibitem[Abdul~Jalil and Wong Ei~Leen(2021)]%
        {abdul_jalil_learning_2021}
\bibfield{author}{\bibinfo{person}{Nasir Abdul~Jalil} {and}
  \bibinfo{person}{Mikkay Wong Ei~Leen}.} \bibinfo{year}{2021}\natexlab{}.
\newblock \showarticletitle{Learning {Analytics} in {Higher} {Education}: {The}
  {Student} {Expectations} of {Learning} {Analytics}}. In
  \bibinfo{booktitle}{\emph{2021 5th {International} {Conference} on
  {Education} and {E}-{Learning}}} \emph{(\bibinfo{series}{{ICEEL} 2021})}.
  \bibinfo{publisher}{Association for Computing Machinery},
  \bibinfo{address}{New York, NY, USA}, \bibinfo{pages}{249--254}.
\newblock
\showISBNx{978-1-4503-8574-9}
\urldef\tempurl%
\url{https://doi.org/10.1145/3502434.3502463}
\showDOI{\tempurl}


\bibitem[Abebe et~al\mbox{.}(2020)]%
        {abebe_roles_2020}
\bibfield{author}{\bibinfo{person}{Rediet Abebe}, \bibinfo{person}{Solon
  Barocas}, \bibinfo{person}{Jon Kleinberg}, \bibinfo{person}{Karen Levy},
  \bibinfo{person}{Manish Raghavan}, {and} \bibinfo{person}{David~G.
  Robinson}.} \bibinfo{year}{2020}\natexlab{}.
\newblock \showarticletitle{Roles for computing in social change}. In
  \bibinfo{booktitle}{\emph{Proceedings of the 2020 {Conference} on {Fairness},
  {Accountability}, and {Transparency}}}. \bibinfo{publisher}{ACM},
  \bibinfo{address}{Barcelona Spain}, \bibinfo{pages}{252--260}.
\newblock
\showISBNx{978-1-4503-6936-7}
\urldef\tempurl%
\url{https://doi.org/10.1145/3351095.3372871}
\showDOI{\tempurl}


\bibitem[Aguiar et~al\mbox{.}(2014)]%
        {aguiar_engagement_2014}
\bibfield{author}{\bibinfo{person}{Everaldo Aguiar}, \bibinfo{person}{Nitesh~V.
  Chawla}, \bibinfo{person}{Jay Brockman}, \bibinfo{person}{G.~Alex Ambrose},
  {and} \bibinfo{person}{Victoria Goodrich}.} \bibinfo{year}{2014}\natexlab{}.
\newblock \showarticletitle{Engagement vs performance: using electronic
  portfolios to predict first semester engineering student retention}. In
  \bibinfo{booktitle}{\emph{Proceedings of the {Fourth} {International}
  {Conference} on {Learning} {Analytics} {And} {Knowledge}}}.
  \bibinfo{publisher}{ACM}, \bibinfo{address}{Indianapolis Indiana USA},
  \bibinfo{pages}{103--112}.
\newblock
\showISBNx{978-1-4503-2664-3}
\urldef\tempurl%
\url{https://doi.org/10.1145/2567574.2567583}
\showDOI{\tempurl}


\bibitem[Ajoodha et~al\mbox{.}(2020)]%
        {ajoodha_forecasting_2020}
\bibfield{author}{\bibinfo{person}{Ritesh Ajoodha}, \bibinfo{person}{Ashwini
  Jadhav}, {and} \bibinfo{person}{Shalini Dukhan}.}
  \bibinfo{year}{2020}\natexlab{}.
\newblock \showarticletitle{Forecasting {Learner} {Attrition} for {Student}
  {Success} at a {South} {African} {University}}. In
  \bibinfo{booktitle}{\emph{Conference of the {South} {African} {Institute} of
  {Computer} {Scientists} and {Information} {Technologists} 2020}}.
  \bibinfo{publisher}{ACM}, \bibinfo{address}{Cape Town South Africa},
  \bibinfo{pages}{19--28}.
\newblock
\showISBNx{978-1-4503-8847-4}
\urldef\tempurl%
\url{https://doi.org/10.1145/3410886.3410973}
\showDOI{\tempurl}


\bibitem[Albreiki et~al\mbox{.}(2021)]%
        {albreiki_systematic_2021}
\bibfield{author}{\bibinfo{person}{Balqis Albreiki}, \bibinfo{person}{Nazar
  Zaki}, {and} \bibinfo{person}{Hany Alashwal}.}
  \bibinfo{year}{2021}\natexlab{}.
\newblock \showarticletitle{A {Systematic} {Literature} {Review} of
  {Student}’ {Performance} {Prediction} {Using} {Machine} {Learning}
  {Techniques}}.
\newblock \bibinfo{journal}{\emph{Education Sciences}} \bibinfo{volume}{11},
  \bibinfo{number}{9} (\bibinfo{date}{Sept.} \bibinfo{year}{2021}),
  \bibinfo{pages}{552}.
\newblock
\showISSN{2227-7102}
\urldef\tempurl%
\url{https://doi.org/10.3390/educsci11090552}
\showDOI{\tempurl}
\newblock
\shownote{Number: 9 Publisher: Multidisciplinary Digital Publishing Institute}.


\bibitem[{Alejandro Jaimes} et~al\mbox{.}(2007)]%
        {alejandro_jaimes_human-centered_2007}
\bibfield{author}{\bibinfo{person}{{Alejandro Jaimes}},
  \bibinfo{person}{{Daniel Gatica-Perez}}, \bibinfo{person}{{Nicu Sebe}}, {and}
  \bibinfo{person}{{Thomas S Huang}}.} \bibinfo{year}{2007}\natexlab{}.
\newblock \showarticletitle{Human-{Centered} {Computing}: {Toward} a {Human}
  {Revolution}}.
\newblock \bibinfo{journal}{\emph{Computer (Long Beach, Calif.)}}
  \bibinfo{volume}{40}, \bibinfo{number}{5} (\bibinfo{year}{2007}),
  \bibinfo{pages}{30--34}.
\newblock
\showISSN{0018-9162}
\urldef\tempurl%
\url{https://search.proquest.com/docview/197419098?pq-origsite=primo}
\showURL{%
\tempurl}
\newblock
\shownote{Place: New York Publisher: The Institute of Electrical and
  Electronics Engineers, Inc. IEEE}.


\bibitem[AlGhamdi et~al\mbox{.}(2020)]%
        {alghamdi_machine_2020}
\bibfield{author}{\bibinfo{person}{Amal AlGhamdi}, \bibinfo{person}{Amal
  Barsheed}, \bibinfo{person}{Hanadi AlMshjary}, {and} \bibinfo{person}{Hanan
  AlGhamdi}.} \bibinfo{year}{2020}\natexlab{}.
\newblock \showarticletitle{A {Machine} {Learning} {Approach} for {Graduate}
  {Admission} {Prediction}}. In \bibinfo{booktitle}{\emph{Proceedings of the
  2020 2nd {International} {Conference} on {Image}, {Video} and {Signal}
  {Processing}}}. \bibinfo{publisher}{ACM}, \bibinfo{address}{Singapore
  Singapore}, \bibinfo{pages}{155--158}.
\newblock
\showISBNx{978-1-4503-7695-2}
\urldef\tempurl%
\url{https://doi.org/10.1145/3388818.3393716}
\showDOI{\tempurl}


\bibitem[Altaf et~al\mbox{.}(2019)]%
        {altaf_student_2019}
\bibfield{author}{\bibinfo{person}{Saud Altaf}, \bibinfo{person}{Waseem
  Soomro}, {and} \bibinfo{person}{Mohd Izani~Mohamed Rawi}.}
  \bibinfo{year}{2019}\natexlab{}.
\newblock \showarticletitle{Student {Performance} {Prediction} using
  {Multi}-{Layers} {Artificial} {Neural} {Networks}: {A} {Case} {Study} on
  {Educational} {Data} {Mining}}. In \bibinfo{booktitle}{\emph{Proceedings of
  the 2019 3rd {International} {Conference} on {Information} {System} and
  {Data} {Mining} - {ICISDM} 2019}}. \bibinfo{publisher}{ACM Press},
  \bibinfo{address}{Houston, TX, USA}, \bibinfo{pages}{59--64}.
\newblock
\showISBNx{978-1-4503-6635-9}
\urldef\tempurl%
\url{https://doi.org/10.1145/3325917.3325919}
\showDOI{\tempurl}


\bibitem[Alyahyan and Düştegör(2020)]%
        {alyahyan_predicting_2020}
\bibfield{author}{\bibinfo{person}{Eyman Alyahyan} {and} \bibinfo{person}{Dilek
  Düştegör}.} \bibinfo{year}{2020}\natexlab{}.
\newblock \showarticletitle{Predicting academic success in higher education:
  literature review and best practices}.
\newblock \bibinfo{journal}{\emph{International Journal of Educational
  Technology in Higher Education}} \bibinfo{volume}{17}, \bibinfo{number}{1}
  (\bibinfo{date}{Dec.} \bibinfo{year}{2020}), \bibinfo{pages}{3}.
\newblock
\showISSN{2365-9440}
\urldef\tempurl%
\url{https://doi.org/10.1186/s41239-020-0177-7}
\showDOI{\tempurl}


\bibitem[Amballoor and Naik(2021)]%
        {amballoor_technological_2021}
\bibfield{author}{\bibinfo{person}{Renji~George Amballoor} {and}
  \bibinfo{person}{Shankar~B Naik}.} \bibinfo{year}{2021}\natexlab{}.
\newblock \showarticletitle{Technological {Achievement} {Index} ({TAI}) in
  {Higher} {Education}- {An} {Empirical} {Analysis}}. In
  \bibinfo{booktitle}{\emph{2021 5th {International} {Conference} on {Digital}
  {Technology} in {Education}}}. \bibinfo{publisher}{ACM},
  \bibinfo{address}{Busan Republic of Korea}, \bibinfo{pages}{93--96}.
\newblock
\showISBNx{978-1-4503-8499-5}
\urldef\tempurl%
\url{https://doi.org/10.1145/3488466.3488476}
\showDOI{\tempurl}


\bibitem[Ameri et~al\mbox{.}(2016)]%
        {ameri_survival_2016}
\bibfield{author}{\bibinfo{person}{Sattar Ameri}, \bibinfo{person}{Mahtab~J.
  Fard}, \bibinfo{person}{Ratna~B. Chinnam}, {and} \bibinfo{person}{Chandan~K.
  Reddy}.} \bibinfo{year}{2016}\natexlab{}.
\newblock \showarticletitle{Survival {Analysis} based {Framework} for {Early}
  {Prediction} of {Student} {Dropouts}}. In
  \bibinfo{booktitle}{\emph{Proceedings of the 25th {ACM} {International} on
  {Conference} on {Information} and {Knowledge} {Management}}}.
  \bibinfo{publisher}{ACM}, \bibinfo{address}{Indianapolis Indiana USA},
  \bibinfo{pages}{903--912}.
\newblock
\showISBNx{978-1-4503-4073-1}
\urldef\tempurl%
\url{https://doi.org/10.1145/2983323.2983351}
\showDOI{\tempurl}


\bibitem[Aragon(2022)]%
        {aragon_human-centered_2022}
\bibfield{author}{\bibinfo{person}{Cecilia~Rodriguez Aragon}.}
  \bibinfo{year}{2022}\natexlab{}.
\newblock \bibinfo{booktitle}{\emph{Human-centered data science: an
  introduction}}.
\newblock \bibinfo{publisher}{The MIT Press}, \bibinfo{address}{Cambridge,
  Massachusetts}.
\newblock
\showISBNx{978-0-262-54321-7}


\bibitem[Bajpai and Mani(2017)]%
        {bajpai_big_2017}
\bibfield{author}{\bibinfo{person}{Swati Bajpai} {and} \bibinfo{person}{S.
  Mani}.} \bibinfo{year}{2017}\natexlab{}.
\newblock \showarticletitle{Big {Data} in {Education} and {Learning}
  {Analytics}}.
\newblock \bibinfo{journal}{\emph{TechnoLearn: An International Journal of
  Educational Technology}} \bibinfo{volume}{7}, \bibinfo{number}{1and2}
  (\bibinfo{year}{2017}), \bibinfo{pages}{45}.
\newblock
\showISSN{2231-4105, 2249-5223}
\urldef\tempurl%
\url{https://doi.org/10.5958/2249-5223.2017.00005.5}
\showDOI{\tempurl}


\bibitem[Baranyi et~al\mbox{.}(2020)]%
        {baranyi_interpretable_2020}
\bibfield{author}{\bibinfo{person}{Máté Baranyi}, \bibinfo{person}{Marcell
  Nagy}, {and} \bibinfo{person}{Roland Molontay}.}
  \bibinfo{year}{2020}\natexlab{}.
\newblock \showarticletitle{Interpretable {Deep} {Learning} for {University}
  {Dropout} {Prediction}}. In \bibinfo{booktitle}{\emph{Proceedings of the 21st
  {Annual} {Conference} on {Information} {Technology} {Education}}}.
  \bibinfo{publisher}{ACM}, \bibinfo{address}{Virtual Event USA},
  \bibinfo{pages}{13--19}.
\newblock
\showISBNx{978-1-4503-7045-5}
\urldef\tempurl%
\url{https://doi.org/10.1145/3368308.3415382}
\showDOI{\tempurl}


\bibitem[Barber and Sharkey(2012)]%
        {barber_course_2012}
\bibfield{author}{\bibinfo{person}{Rebecca Barber} {and} \bibinfo{person}{Mike
  Sharkey}.} \bibinfo{year}{2012}\natexlab{}.
\newblock \showarticletitle{Course correction: using analytics to predict
  course success}. In \bibinfo{booktitle}{\emph{Proceedings of the 2nd
  {International} {Conference} on {Learning} {Analytics} and {Knowledge}}}.
  \bibinfo{publisher}{ACM}, \bibinfo{address}{Vancouver British Columbia
  Canada}, \bibinfo{pages}{259--262}.
\newblock
\showISBNx{978-1-4503-1111-3}
\urldef\tempurl%
\url{https://doi.org/10.1145/2330601.2330664}
\showDOI{\tempurl}


\bibitem[Barocas(2014)]%
        {barocas_putting_2014}
\bibfield{author}{\bibinfo{person}{Solon Barocas}.}
  \bibinfo{year}{2014}\natexlab{}.
\newblock \showarticletitle{Putting {Data} to {Work}}.
\newblock In \bibinfo{booktitle}{\emph{Data and {Discrimination}: {Collected}
  {Essays}}}, \bibfield{editor}{\bibinfo{person}{Seeta Peña~Gangadharan},
  \bibinfo{person}{Virginia Eubanks}, {and} \bibinfo{person}{Solon Barocas}}
  (Eds.). \bibinfo{publisher}{Open Technology Institute}.
\newblock
\urldef\tempurl%
\url{https://timlibert.me/pdf/2014-Data_Discrimination_Collected_Essays.pdf}
\showURL{%
\tempurl}


\bibitem[Baumer(2017)]%
        {baumer_toward_2017}
\bibfield{author}{\bibinfo{person}{Eric~PS Baumer}.}
  \bibinfo{year}{2017}\natexlab{}.
\newblock \showarticletitle{Toward human-centered algorithm design}.
\newblock \bibinfo{journal}{\emph{Big Data \& Society}} \bibinfo{volume}{4},
  \bibinfo{number}{2} (\bibinfo{date}{Dec.} \bibinfo{year}{2017}),
  \bibinfo{pages}{2053951717718854}.
\newblock
\showISSN{2053-9517}
\urldef\tempurl%
\url{https://doi.org/10.1177/2053951717718854}
\showDOI{\tempurl}
\newblock
\shownote{Publisher: SAGE Publications Ltd}.


\bibitem[Benablo et~al\mbox{.}(2018)]%
        {benablo_higher_2018}
\bibfield{author}{\bibinfo{person}{Ceasar Ian~P. Benablo},
  \bibinfo{person}{Evangeline~T. Sarte}, \bibinfo{person}{Joe Marie~D.
  Dormido}, {and} \bibinfo{person}{Thelma Palaoag}.}
  \bibinfo{year}{2018}\natexlab{}.
\newblock \showarticletitle{Higher {Education} {Student}'s {Academic}
  {Performance} {Analysis} through {Predictive} {Analytics}}. In
  \bibinfo{booktitle}{\emph{Proceedings of the 2018 7th {International}
  {Conference} on {Software} and {Computer} {Applications}}}.
  \bibinfo{publisher}{ACM}, \bibinfo{address}{Kuantan Malaysia},
  \bibinfo{pages}{238--242}.
\newblock
\showISBNx{978-1-4503-5414-1}
\urldef\tempurl%
\url{https://doi.org/10.1145/3185089.3185102}
\showDOI{\tempurl}


\bibitem[Borrella et~al\mbox{.}(2019)]%
        {borrella_predict_2019}
\bibfield{author}{\bibinfo{person}{Inma Borrella}, \bibinfo{person}{Sergio
  Caballero-Caballero}, {and} \bibinfo{person}{Eva Ponce-Cueto}.}
  \bibinfo{year}{2019}\natexlab{}.
\newblock \showarticletitle{Predict and {Intervene}: {Addressing} the {Dropout}
  {Problem} in a {MOOC}-based {Program}}. In
  \bibinfo{booktitle}{\emph{Proceedings of the {Sixth} (2019) {ACM}
  {Conference} on {Learning} @ {Scale}}}. \bibinfo{publisher}{ACM},
  \bibinfo{address}{Chicago IL USA}, \bibinfo{pages}{1--9}.
\newblock
\showISBNx{978-1-4503-6804-9}
\urldef\tempurl%
\url{https://doi.org/10.1145/3330430.3333634}
\showDOI{\tempurl}


\bibitem[Bos and Brand-Gruwel(2016)]%
        {bos_student_2016}
\bibfield{author}{\bibinfo{person}{Nynke Bos} {and} \bibinfo{person}{Saskia
  Brand-Gruwel}.} \bibinfo{year}{2016}\natexlab{}.
\newblock \showarticletitle{Student differences in regulation strategies and
  their use of learning resources: implications for educational design}. In
  \bibinfo{booktitle}{\emph{Proceedings of the {Sixth} {International}
  {Conference} on {Learning} {Analytics} \& {Knowledge} - {LAK} '16}}.
  \bibinfo{publisher}{ACM Press}, \bibinfo{address}{Edinburgh, United Kingdom},
  \bibinfo{pages}{344--353}.
\newblock
\showISBNx{978-1-4503-4190-5}
\urldef\tempurl%
\url{https://doi.org/10.1145/2883851.2883890}
\showDOI{\tempurl}


\bibitem[Brooks and Greer(2014)]%
        {brooks_explaining_2014}
\bibfield{author}{\bibinfo{person}{Christopher Brooks} {and}
  \bibinfo{person}{Jim Greer}.} \bibinfo{year}{2014}\natexlab{}.
\newblock \showarticletitle{Explaining predictive models to learning
  specialists using personas}. In \bibinfo{booktitle}{\emph{Proceedings of the
  {Fourth} {International} {Conference} on {Learning} {Analytics} {And}
  {Knowledge}}} \emph{(\bibinfo{series}{{LAK} '14})}.
  \bibinfo{publisher}{Association for Computing Machinery},
  \bibinfo{address}{New York, NY, USA}, \bibinfo{pages}{26--30}.
\newblock
\showISBNx{978-1-4503-2664-3}
\urldef\tempurl%
\url{https://doi.org/10.1145/2567574.2567612}
\showDOI{\tempurl}


\bibitem[Buerck et~al\mbox{.}(2013)]%
        {buerck_predicting_2013}
\bibfield{author}{\bibinfo{person}{John~P. Buerck},
  \bibinfo{person}{Srikanth~P. Mudigonda}, \bibinfo{person}{Stephanie~E.
  Mooshegian}, \bibinfo{person}{Kyle Collins}, \bibinfo{person}{Nicholas
  Grimm}, \bibinfo{person}{Kristen Bonney}, {and} \bibinfo{person}{Hadley
  Kombrink}.} \bibinfo{year}{2013}\natexlab{}.
\newblock \showarticletitle{Predicting non-traditional student learning
  outcomes using data analytics - a pilot research study}.
\newblock \bibinfo{journal}{\emph{Journal of Computing Sciences in Colleges}}
  \bibinfo{volume}{28}, \bibinfo{number}{5} (\bibinfo{date}{May}
  \bibinfo{year}{2013}), \bibinfo{pages}{260--265}.
\newblock
\showISSN{1937-4771}


\bibitem[Cabrera et~al\mbox{.}(2020)]%
        {cabrera_data_2020}
\bibfield{author}{\bibinfo{person}{Joey~A. Cabrera}, \bibinfo{person}{Markdy~Y.
  Orong}, \bibinfo{person}{Nelpa~N. Capio}, \bibinfo{person}{Arnel Filarca},
  \bibinfo{person}{Eden Neri}, {and} \bibinfo{person}{Ariel~R. Clarin}.}
  \bibinfo{year}{2020}\natexlab{}.
\newblock \showarticletitle{A {Data} {Mining} {Approach} for {Student}
  {Referral} {Service} of the {Guidance} {Center}: {An} {Input} in {Designing}
  {Mediation} {Scheme} for {Higher} {Education} {Institutions} of the
  {Philippines}}. In \bibinfo{booktitle}{\emph{Proceedings of the 3rd
  {International} {Conference} on {Software} {Engineering} and {Information}
  {Management}}}. \bibinfo{publisher}{ACM}, \bibinfo{address}{Sydney NSW
  Australia}, \bibinfo{pages}{10--14}.
\newblock
\showISBNx{978-1-4503-7690-7}
\urldef\tempurl%
\url{https://doi.org/10.1145/3378936.3378958}
\showDOI{\tempurl}


\bibitem[Campbell and Dickson(1996)]%
        {campbell_predicting_1996}
\bibfield{author}{\bibinfo{person}{Arthur~Ree Campbell} {and}
  \bibinfo{person}{Charlie~J. Dickson}.} \bibinfo{year}{1996}\natexlab{}.
\newblock \showarticletitle{Predicting {Student} {Success}: {A} 10-{Year}
  {Review} {Using} {Integrative} {Review} and {Meta}-{Analysis}}.
\newblock \bibinfo{journal}{\emph{Journal of Professional Nursing}}
  \bibinfo{volume}{12}, \bibinfo{number}{1} (\bibinfo{year}{1996}),
  \bibinfo{pages}{47--59}.
\newblock
\showISSN{8755-7223}


\bibitem[Chancellor et~al\mbox{.}(2019)]%
        {chancellor_who_2019}
\bibfield{author}{\bibinfo{person}{Stevie Chancellor}, \bibinfo{person}{Eric
  P.~S. Baumer}, {and} \bibinfo{person}{Munmun De~Choudhury}.}
  \bibinfo{year}{2019}\natexlab{}.
\newblock \showarticletitle{Who is the "{Human}" in {Human}-{Centered}
  {Machine} {Learning}: {The} {Case} of {Predicting} {Mental} {Health} from
  {Social} {Media}}.
\newblock \bibinfo{journal}{\emph{Proceedings of the ACM on Human-Computer
  Interaction}} \bibinfo{volume}{3}, \bibinfo{number}{CSCW}
  (\bibinfo{date}{Nov.} \bibinfo{year}{2019}), \bibinfo{pages}{1--32}.
\newblock
\showISSN{2573-0142}
\urldef\tempurl%
\url{https://doi.org/10.1145/3359249}
\showDOI{\tempurl}


\bibitem[Chen et~al\mbox{.}(2022)]%
        {chen_pathways_2022}
\bibfield{author}{\bibinfo{person}{Youjie Chen}, \bibinfo{person}{Annie Fu},
  \bibinfo{person}{Jennifer Jia-Ling Lee}, \bibinfo{person}{Ian~Wilkie
  Tomasik}, {and} \bibinfo{person}{René~F. Kizilcec}.}
  \bibinfo{year}{2022}\natexlab{}.
\newblock \showarticletitle{Pathways: {Exploring} {Academic} {Interests} with
  {Historical} {Course} {Enrollment} {Records}}. In
  \bibinfo{booktitle}{\emph{Proceedings of the {Ninth} {ACM} {Conference} on
  {Learning} @ {Scale}}}. \bibinfo{publisher}{ACM}, \bibinfo{address}{New York
  City NY USA}, \bibinfo{pages}{222--233}.
\newblock
\showISBNx{978-1-4503-9158-0}
\urldef\tempurl%
\url{https://doi.org/10.1145/3491140.3528270}
\showDOI{\tempurl}


\bibitem[Cherrington et~al\mbox{.}(2020)]%
        {cherrington_features_2020}
\bibfield{author}{\bibinfo{person}{Marianne Cherrington},
  \bibinfo{person}{David Airehrour}, \bibinfo{person}{Joan Lu},
  \bibinfo{person}{Qiang Xu}, \bibinfo{person}{David Cameron-Brown}, {and}
  \bibinfo{person}{Ihaka Dunn}.} \bibinfo{year}{2020}\natexlab{}.
\newblock \showarticletitle{Features of {Human}-{Centred} {Algorithm}
  {Design}}. In \bibinfo{booktitle}{\emph{2020 30th {International}
  {Telecommunication} {Networks} and {Applications} {Conference} ({ITNAC})}}.
  \bibinfo{pages}{1--6}.
\newblock
\urldef\tempurl%
\url{https://doi.org/10.1109/ITNAC50341.2020.9315169}
\showDOI{\tempurl}
\newblock
\shownote{ISSN: 2474-154X}.


\bibitem[Chockkalingam et~al\mbox{.}(2021)]%
        {chockkalingam_which_2021}
\bibfield{author}{\bibinfo{person}{Shruthi Chockkalingam}, \bibinfo{person}{Run
  Yu}, {and} \bibinfo{person}{Zachary~A. Pardos}.}
  \bibinfo{year}{2021}\natexlab{}.
\newblock \showarticletitle{Which one’s more work? {Predicting} effective
  credit hours between courses}. In \bibinfo{booktitle}{\emph{{LAK21}: 11th
  {International} {Learning} {Analytics} and {Knowledge} {Conference}}}.
  \bibinfo{publisher}{ACM}, \bibinfo{address}{Irvine CA USA},
  \bibinfo{pages}{599--605}.
\newblock
\showISBNx{978-1-4503-8935-8}
\urldef\tempurl%
\url{https://doi.org/10.1145/3448139.3448204}
\showDOI{\tempurl}


\bibitem[Chong et~al\mbox{.}(2020)]%
        {chong_data_2020}
\bibfield{author}{\bibinfo{person}{Sylvia Chong}, \bibinfo{person}{Yew~Haur
  Lee}, {and} \bibinfo{person}{Yoke~Wah Tang}.}
  \bibinfo{year}{2020}\natexlab{}.
\newblock \showarticletitle{Data {Analytics} and {Visualization} to {Support}
  the {Adult} {Learner} in {Higher} {Education}}. In
  \bibinfo{booktitle}{\emph{2020 {The} 4th {International} {Conference} on
  {E}-{Society}, {E}-{Education} and {E}-{Technology}}}.
  \bibinfo{publisher}{ACM}, \bibinfo{address}{Taipei Taiwan},
  \bibinfo{pages}{126--131}.
\newblock
\showISBNx{978-1-4503-8877-1}
\urldef\tempurl%
\url{https://doi.org/10.1145/3421682.3421698}
\showDOI{\tempurl}


\bibitem[Daud et~al\mbox{.}(2017)]%
        {daud_predicting_2017}
\bibfield{author}{\bibinfo{person}{Ali Daud}, \bibinfo{person}{Naif~Radi
  Aljohani}, \bibinfo{person}{Rabeeh~Ayaz Abbasi},
  \bibinfo{person}{Miltiadis~D. Lytras}, \bibinfo{person}{Farhat Abbas}, {and}
  \bibinfo{person}{Jalal~S. Alowibdi}.} \bibinfo{year}{2017}\natexlab{}.
\newblock \showarticletitle{Predicting {Student} {Performance} using {Advanced}
  {Learning} {Analytics}}. In \bibinfo{booktitle}{\emph{Proceedings of the 26th
  {International} {Conference} on {World} {Wide} {Web} {Companion} - {WWW} '17
  {Companion}}}. \bibinfo{publisher}{ACM Press}, \bibinfo{address}{Perth,
  Australia}, \bibinfo{pages}{415--421}.
\newblock
\showISBNx{978-1-4503-4914-7}
\urldef\tempurl%
\url{https://doi.org/10.1145/3041021.3054164}
\showDOI{\tempurl}


\bibitem[Dawson et~al\mbox{.}(2017)]%
        {dawson_prediction_2017}
\bibfield{author}{\bibinfo{person}{Shane Dawson}, \bibinfo{person}{Jelena
  Jovanovic}, \bibinfo{person}{Dragan Gašević}, {and}
  \bibinfo{person}{Abelardo Pardo}.} \bibinfo{year}{2017}\natexlab{}.
\newblock \showarticletitle{From prediction to impact: evaluation of a learning
  analytics retention program}. In \bibinfo{booktitle}{\emph{Proceedings of the
  {Seventh} {International} {Learning} {Analytics} \& {Knowledge}
  {Conference}}}. \bibinfo{publisher}{ACM}, \bibinfo{address}{Vancouver British
  Columbia Canada}, \bibinfo{pages}{474--478}.
\newblock
\showISBNx{978-1-4503-4870-6}
\urldef\tempurl%
\url{https://doi.org/10.1145/3027385.3027405}
\showDOI{\tempurl}


\bibitem[de~Brey et~al\mbox{.}(2019)]%
        {de_brey_status_2019}
\bibfield{author}{\bibinfo{person}{Cristobal de Brey}, \bibinfo{person}{Lauren
  Musu}, \bibinfo{person}{Joel McFarland}, \bibinfo{person}{Sidney
  Wilkinson-Flicker}, \bibinfo{person}{Melissa Diliberti},
  \bibinfo{person}{Anlan Zhang}, \bibinfo{person}{Claire Branstetter}, {and}
  \bibinfo{person}{Xiaolei Wang}.} \bibinfo{year}{2019}\natexlab{}.
\newblock \bibinfo{booktitle}{\emph{Status and {Trends} in the {Education} of
  {Racial} and {Ethnic} {Groups} 2018}}.
\newblock \bibinfo{type}{{T}echnical {R}eport}. \bibinfo{institution}{U.S.
  Department of Education, National Center for Education Statistics, Institute
  of Education Sciences}. \bibinfo{pages}{228} pages.
\newblock


\bibitem[Delgado et~al\mbox{.}(2022)]%
        {delgado_uncommon_2022}
\bibfield{author}{\bibinfo{person}{Fernando Delgado}, \bibinfo{person}{Solon
  Barocas}, {and} \bibinfo{person}{Karen Levy}.}
  \bibinfo{year}{2022}\natexlab{}.
\newblock \showarticletitle{An {Uncommon} {Task}: {Participatory} {Design} in
  {Legal} {AI}}.
\newblock \bibinfo{journal}{\emph{Proceedings of the ACM on Human-Computer
  Interaction}} \bibinfo{volume}{6}, \bibinfo{number}{CSCW1}
  (\bibinfo{date}{April} \bibinfo{year}{2022}), \bibinfo{pages}{51:1--51:23}.
\newblock
\urldef\tempurl%
\url{https://doi.org/10.1145/3512898}
\showDOI{\tempurl}


\bibitem[Demmans~Epp et~al\mbox{.}(2020)]%
        {demmans_epp_learning_2020}
\bibfield{author}{\bibinfo{person}{Carrie Demmans~Epp},
  \bibinfo{person}{Krystle Phirangee}, \bibinfo{person}{Jim Hewitt}, {and}
  \bibinfo{person}{Charles~A. Perfetti}.} \bibinfo{year}{2020}\natexlab{}.
\newblock \showarticletitle{Learning management system and course influences on
  student actions and learning experiences}.
\newblock \bibinfo{journal}{\emph{Educational Technology Research and
  Development}} \bibinfo{volume}{68}, \bibinfo{number}{6} (\bibinfo{date}{Dec.}
  \bibinfo{year}{2020}), \bibinfo{pages}{3263--3297}.
\newblock
\showISSN{1042-1629, 1556-6501}
\urldef\tempurl%
\url{https://doi.org/10.1007/s11423-020-09821-1}
\showDOI{\tempurl}


\bibitem[Echeverría et~al\mbox{.}(2014)]%
        {echeverria_presentation_2014}
\bibfield{author}{\bibinfo{person}{Vanessa Echeverría}, \bibinfo{person}{Allan
  Avendaño}, \bibinfo{person}{Katherine Chiluiza}, \bibinfo{person}{Aníbal
  Vásquez}, {and} \bibinfo{person}{Xavier Ochoa}.}
  \bibinfo{year}{2014}\natexlab{}.
\newblock \showarticletitle{Presentation {Skills} {Estimation} {Based} on
  {Video} and {Kinect} {Data} {Analysis}}. In
  \bibinfo{booktitle}{\emph{Proceedings of the 2014 {ACM} workshop on
  {Multimodal} {Learning} {Analytics} {Workshop} and {Grand} {Challenge}}}.
  \bibinfo{publisher}{ACM}, \bibinfo{address}{Istanbul Turkey},
  \bibinfo{pages}{53--60}.
\newblock
\showISBNx{978-1-4503-0488-7}
\urldef\tempurl%
\url{https://doi.org/10.1145/2666633.2666641}
\showDOI{\tempurl}


\bibitem[Edwards et~al\mbox{.}(2017)]%
        {edwards_using_2017}
\bibfield{author}{\bibinfo{person}{Rebecca~L. Edwards},
  \bibinfo{person}{Sarah~K. Davis}, \bibinfo{person}{Allyson~F. Hadwin}, {and}
  \bibinfo{person}{Todd~M. Milford}.} \bibinfo{year}{2017}\natexlab{}.
\newblock \showarticletitle{Using predictive analytics in a self-regulated
  learning university course to promote student success}. In
  \bibinfo{booktitle}{\emph{Proceedings of the {Seventh} {International}
  {Learning} {Analytics} \& {Knowledge} {Conference}}}.
  \bibinfo{publisher}{ACM}, \bibinfo{address}{Vancouver British Columbia
  Canada}, \bibinfo{pages}{556--557}.
\newblock
\showISBNx{978-1-4503-4870-6}
\urldef\tempurl%
\url{https://doi.org/10.1145/3027385.3029455}
\showDOI{\tempurl}


\bibitem[Errahmouni~Barkam et~al\mbox{.}(2022)]%
        {errahmouni_barkam_testing_2022}
\bibfield{author}{\bibinfo{person}{Hamza Errahmouni~Barkam},
  \bibinfo{person}{Max Wang}, \bibinfo{person}{Barbara Martinez~Neda}, {and}
  \bibinfo{person}{Sergio Gago-Masague}.} \bibinfo{year}{2022}\natexlab{}.
\newblock \showarticletitle{Testing {Machine} {Learning} {Models} to {Identify}
  {Computer} {Science} {Students} at {High}-risk of {Probation}}. In
  \bibinfo{booktitle}{\emph{Proceedings of the 53rd {ACM} {Technical}
  {Symposium} on {Computer} {Science} {Education} {V}. 2}}.
  \bibinfo{publisher}{ACM}, \bibinfo{address}{Providence RI USA},
  \bibinfo{pages}{1161--1161}.
\newblock
\showISBNx{978-1-4503-9071-2}
\urldef\tempurl%
\url{https://doi.org/10.1145/3478432.3499103}
\showDOI{\tempurl}


\bibitem[Feild et~al\mbox{.}(2018)]%
        {feild_generalized_2018}
\bibfield{author}{\bibinfo{person}{Jacqueline Feild}, \bibinfo{person}{Nicholas
  Lewkow}, \bibinfo{person}{Sean Burns}, {and} \bibinfo{person}{Karen
  Gebhardt}.} \bibinfo{year}{2018}\natexlab{}.
\newblock \showarticletitle{A generalized classifier to identify online
  learning tool disengagement at scale}. In
  \bibinfo{booktitle}{\emph{Proceedings of the 8th {International} {Conference}
  on {Learning} {Analytics} and {Knowledge}}}. \bibinfo{publisher}{ACM},
  \bibinfo{address}{Sydney New South Wales Australia}, \bibinfo{pages}{61--70}.
\newblock
\showISBNx{978-1-4503-6400-3}
\urldef\tempurl%
\url{https://doi.org/10.1145/3170358.3170370}
\showDOI{\tempurl}


\bibitem[Fiorini et~al\mbox{.}(2018)]%
        {fiorini_application_2018}
\bibfield{author}{\bibinfo{person}{Stefano Fiorini}, \bibinfo{person}{Adrienne
  Sewell}, \bibinfo{person}{Mathew Bumbalough}, \bibinfo{person}{Pallavi
  Chauhan}, \bibinfo{person}{Linda Shepard}, \bibinfo{person}{George Rehrey},
  {and} \bibinfo{person}{Dennis Groth}.} \bibinfo{year}{2018}\natexlab{}.
\newblock \showarticletitle{An application of participatory action research in
  advising-focused learning analytics}. In
  \bibinfo{booktitle}{\emph{Proceedings of the 8th {International} {Conference}
  on {Learning} {Analytics} and {Knowledge}}}. \bibinfo{publisher}{ACM},
  \bibinfo{address}{Sydney New South Wales Australia}, \bibinfo{pages}{89--96}.
\newblock
\showISBNx{978-1-4503-6400-3}
\urldef\tempurl%
\url{https://doi.org/10.1145/3170358.3170387}
\showDOI{\tempurl}


\bibitem[Flügge(2020)]%
        {flugge_algorithmic_2020}
\bibfield{author}{\bibinfo{person}{Asbjørn~Ammitzbøll Flügge}.}
  \bibinfo{year}{2020}\natexlab{}.
\newblock \showarticletitle{Algorithmic {Decision} {Making} in {Public}
  {Administration}: {A} {CSCW}-{Perspective}}. In
  \bibinfo{booktitle}{\emph{Companion of the 2020 {ACM} {International}
  {Conference} on {Supporting} {Group} {Work}}} \emph{(\bibinfo{series}{{GROUP}
  '20})}. \bibinfo{publisher}{Association for Computing Machinery},
  \bibinfo{address}{New York, NY, USA}, \bibinfo{pages}{15--24}.
\newblock
\showISBNx{978-1-4503-6767-7}
\urldef\tempurl%
\url{https://doi.org/10.1145/3323994.3371016}
\showDOI{\tempurl}


\bibitem[Fox et~al\mbox{.}(2020)]%
        {fox_worker-centered_2020}
\bibfield{author}{\bibinfo{person}{Sarah~E. Fox}, \bibinfo{person}{Vera
  Khovanskaya}, \bibinfo{person}{Clara Crivellaro}, \bibinfo{person}{Niloufar
  Salehi}, \bibinfo{person}{Lynn Dombrowski}, \bibinfo{person}{Chinmay
  Kulkarni}, \bibinfo{person}{Lilly Irani}, {and} \bibinfo{person}{Jodi
  Forlizzi}.} \bibinfo{year}{2020}\natexlab{}.
\newblock \showarticletitle{Worker-{Centered} {Design}: {Expanding} {HCI}
  {Methods} for {Supporting} {Labor}}. In \bibinfo{booktitle}{\emph{Extended
  {Abstracts} of the 2020 {CHI} {Conference} on {Human} {Factors} in
  {Computing} {Systems}}} \emph{(\bibinfo{series}{{CHI} {EA} '20})}.
  \bibinfo{publisher}{Association for Computing Machinery},
  \bibinfo{address}{New York, NY, USA}, \bibinfo{pages}{1--8}.
\newblock
\showISBNx{978-1-4503-6819-3}
\urldef\tempurl%
\url{https://doi.org/10.1145/3334480.3375157}
\showDOI{\tempurl}


\bibitem[Gamie et~al\mbox{.}(2019)]%
        {gamie_layered-analysis_2019}
\bibfield{author}{\bibinfo{person}{Eslam~Abou Gamie},
  \bibinfo{person}{M.~Samir~Abou El-Seoud}, {and} \bibinfo{person}{Mostafa~A.
  Salama}.} \bibinfo{year}{2019}\natexlab{}.
\newblock \showarticletitle{A layered-analysis of the features in higher
  education data set}. In \bibinfo{booktitle}{\emph{Proceedings of the 2019 8th
  {International} {Conference} on {Software} and {Information} {Engineering}}}.
  \bibinfo{publisher}{ACM}, \bibinfo{address}{Cairo Egypt},
  \bibinfo{pages}{237--242}.
\newblock
\showISBNx{978-1-4503-6105-7}
\urldef\tempurl%
\url{https://doi.org/10.1145/3328833.3328850}
\showDOI{\tempurl}


\bibitem[Gatbonton and Aguinaldo(2018)]%
        {gatbonton_employability_2018}
\bibfield{author}{\bibinfo{person}{Twinkle Mae~C. Gatbonton} {and}
  \bibinfo{person}{Betchie~E. Aguinaldo}.} \bibinfo{year}{2018}\natexlab{}.
\newblock \showarticletitle{Employability {Predictive} {Model} {Evaluator}
  {Using} {PART} and {JRip} {Classifier}}. In
  \bibinfo{booktitle}{\emph{Proceedings of the 6th {International} {Conference}
  on {Information} {Technology}: {IoT} and {Smart} {City} - {ICIT} 2018}}.
  \bibinfo{publisher}{ACM Press}, \bibinfo{address}{Hong Kong, Hong Kong},
  \bibinfo{pages}{307--310}.
\newblock
\showISBNx{978-1-4503-6629-8}
\urldef\tempurl%
\url{https://doi.org/10.1145/3301551.3301569}
\showDOI{\tempurl}


\bibitem[Guyon and Elisseeff(2003)]%
        {guyon_introduction_2003}
\bibfield{author}{\bibinfo{person}{Isabelle Guyon} {and}
  \bibinfo{person}{André Elisseeff}.} \bibinfo{year}{2003}\natexlab{}.
\newblock \showarticletitle{An introduction to variable and feature selection}.
\newblock \bibinfo{journal}{\emph{The Journal of Machine Learning Research}}
  \bibinfo{volume}{3}, \bibinfo{number}{null} (\bibinfo{date}{March}
  \bibinfo{year}{2003}), \bibinfo{pages}{1157--1182}.
\newblock
\showISSN{1532-4435}


\bibitem[Hall(2012)]%
        {hall_incident_2012}
\bibfield{author}{\bibinfo{person}{Patrick Hall}.}
  \bibinfo{year}{2012}\natexlab{}.
\newblock \bibinfo{title}{Incident 135: {University} of {Texas} at {Austin}’s
  {Algorithm} to {Evaluate} {Graduate} {Applications}, {GRADE}, {Allegedly}
  {Exacerbated} {Existing} {Inequality} for {Marginalized} {Applicants},
  {Prompting} {Tool} {Suspension}}.
\newblock
\newblock
\urldef\tempurl%
\url{https://incidentdatabase.ai/cite/135}
\showURL{%
\tempurl}


\bibitem[Hall(2020)]%
        {hall_incident_2020}
\bibfield{author}{\bibinfo{person}{Patrick Hall}.}
  \bibinfo{year}{2020}\natexlab{}.
\newblock \bibinfo{title}{Incident 140: {ProctorU}’s {Identity}
  {Verification} and {Exam} {Monitoring} {Systems} {Provided} {Allegedly}
  {Discriminatory} {Experiences} for {BIPOC} {Students}}.
\newblock
\newblock
\urldef\tempurl%
\url{https://incidentdatabase.ai/cite/140}
\showURL{%
\tempurl}


\bibitem[Harackiewicz and Priniski(2018)]%
        {harackiewicz_improving_2018}
\bibfield{author}{\bibinfo{person}{Judith~M. Harackiewicz} {and}
  \bibinfo{person}{Stacy~J. Priniski}.} \bibinfo{year}{2018}\natexlab{}.
\newblock \showarticletitle{Improving {Student} {Outcomes} in {Higher}
  {Education}: {The} {Science} of {Targeted} {Intervention}}.
\newblock \bibinfo{journal}{\emph{Annual review of psychology}}
  \bibinfo{volume}{69} (\bibinfo{date}{Jan.} \bibinfo{year}{2018}),
  \bibinfo{pages}{409--435}.
\newblock
\showISSN{0066-4308}
\urldef\tempurl%
\url{https://doi.org/10.1146/annurev-psych-122216-011725}
\showDOI{\tempurl}


\bibitem[Hellas et~al\mbox{.}(2018)]%
        {hellas_predicting_2018}
\bibfield{author}{\bibinfo{person}{Arto Hellas}, \bibinfo{person}{Petri
  Ihantola}, \bibinfo{person}{Andrew Petersen}, \bibinfo{person}{Vangel~V.
  Ajanovski}, \bibinfo{person}{Mirela Gutica}, \bibinfo{person}{Timo Hynninen},
  \bibinfo{person}{Antti Knutas}, \bibinfo{person}{Juho Leinonen},
  \bibinfo{person}{Chris Messom}, {and} \bibinfo{person}{Soohyun~Nam Liao}.}
  \bibinfo{year}{2018}\natexlab{}.
\newblock \showarticletitle{Predicting academic performance: a systematic
  literature review}. In \bibinfo{booktitle}{\emph{Proceedings {Companion} of
  the 23rd {Annual} {ACM} {Conference} on {Innovation} and {Technology} in
  {Computer} {Science} {Education}}}. \bibinfo{publisher}{ACM},
  \bibinfo{address}{Larnaca Cyprus}, \bibinfo{pages}{175--199}.
\newblock
\showISBNx{978-1-4503-6223-8}
\urldef\tempurl%
\url{https://doi.org/10.1145/3293881.3295783}
\showDOI{\tempurl}


\bibitem[Hien et~al\mbox{.}(2018)]%
        {hien_intelligent_2018}
\bibfield{author}{\bibinfo{person}{Ho~Thao Hien}, \bibinfo{person}{Pham-Nguyen
  Cuong}, \bibinfo{person}{Le~Nguyen~Hoai Nam}, \bibinfo{person}{Ho~Le Thi~Kim
  Nhung}, {and} \bibinfo{person}{Le~Dinh Thang}.}
  \bibinfo{year}{2018}\natexlab{}.
\newblock \showarticletitle{Intelligent {Assistants} in {Higher}-{Education}
  {Environments}: {The} {FIT}-{EBot}, a {Chatbot} for {Administrative} and
  {Learning} {Support}}. In \bibinfo{booktitle}{\emph{Proceedings of the
  {Ninth} {International} {Symposium} on {Information} and {Communication}
  {Technology} - {SoICT} 2018}}. \bibinfo{publisher}{ACM Press},
  \bibinfo{address}{Danang City, Viet Nam}, \bibinfo{pages}{69--76}.
\newblock
\showISBNx{978-1-4503-6539-0}
\urldef\tempurl%
\url{https://doi.org/10.1145/3287921.3287937}
\showDOI{\tempurl}


\bibitem[Hlosta et~al\mbox{.}(2017)]%
        {hlosta_ouroboros_2017}
\bibfield{author}{\bibinfo{person}{Martin Hlosta}, \bibinfo{person}{Zdenek
  Zdrahal}, {and} \bibinfo{person}{Jaroslav Zendulka}.}
  \bibinfo{year}{2017}\natexlab{}.
\newblock \showarticletitle{Ouroboros: early identification of at-risk students
  without models based on legacy data}. In
  \bibinfo{booktitle}{\emph{Proceedings of the {Seventh} {International}
  {Learning} {Analytics} \& {Knowledge} {Conference}}}.
  \bibinfo{publisher}{ACM}, \bibinfo{address}{Vancouver British Columbia
  Canada}, \bibinfo{pages}{6--15}.
\newblock
\showISBNx{978-1-4503-4870-6}
\urldef\tempurl%
\url{https://doi.org/10.1145/3027385.3027449}
\showDOI{\tempurl}


\bibitem[Jayaprakash and Lauría(2014)]%
        {jayaprakash_open_2014}
\bibfield{author}{\bibinfo{person}{Sandeep~M. Jayaprakash} {and}
  \bibinfo{person}{Eitel J.~M. Lauría}.} \bibinfo{year}{2014}\natexlab{}.
\newblock \showarticletitle{Open academic early alert system: technical
  demonstration}. In \bibinfo{booktitle}{\emph{Proceedings of the {Fourth}
  {International} {Conference} on {Learning} {Analytics} {And} {Knowledge}}}.
  \bibinfo{publisher}{ACM}, \bibinfo{address}{Indianapolis Indiana USA},
  \bibinfo{pages}{267--268}.
\newblock
\showISBNx{978-1-4503-2664-3}
\urldef\tempurl%
\url{https://doi.org/10.1145/2567574.2567578}
\showDOI{\tempurl}


\bibitem[Jiang and Pardos(2019)]%
        {jiang_time_2019}
\bibfield{author}{\bibinfo{person}{Weijie Jiang} {and}
  \bibinfo{person}{Zachary~A. Pardos}.} \bibinfo{year}{2019}\natexlab{}.
\newblock \showarticletitle{Time slice imputation for personalized goal-based
  recommendation in higher education}. In \bibinfo{booktitle}{\emph{Proceedings
  of the 13th {ACM} {Conference} on {Recommender} {Systems}}}.
  \bibinfo{publisher}{ACM}, \bibinfo{address}{Copenhagen Denmark},
  \bibinfo{pages}{506--510}.
\newblock
\showISBNx{978-1-4503-6243-6}
\urldef\tempurl%
\url{https://doi.org/10.1145/3298689.3347030}
\showDOI{\tempurl}


\bibitem[Jiang and Pardos(2021)]%
        {jiang_towards_2021}
\bibfield{author}{\bibinfo{person}{Weijie Jiang} {and}
  \bibinfo{person}{Zachary~A. Pardos}.} \bibinfo{year}{2021}\natexlab{}.
\newblock \showarticletitle{Towards {Equity} and {Algorithmic} {Fairness} in
  {Student} {Grade} {Prediction}}. In \bibinfo{booktitle}{\emph{Proceedings of
  the 2021 {AAAI}/{ACM} {Conference} on {AI}, {Ethics}, and {Society}}}.
  \bibinfo{publisher}{ACM}, \bibinfo{address}{Virtual Event USA},
  \bibinfo{pages}{608--617}.
\newblock
\showISBNx{978-1-4503-8473-5}
\urldef\tempurl%
\url{https://doi.org/10.1145/3461702.3462623}
\showDOI{\tempurl}


\bibitem[Karypis(2017)]%
        {karypis_improving_2017}
\bibfield{author}{\bibinfo{person}{George Karypis}.}
  \bibinfo{year}{2017}\natexlab{}.
\newblock \showarticletitle{Improving {Higher} {Education}: {Learning}
  {Analytics} \& {Recommender} {Systems} {Research}}. In
  \bibinfo{booktitle}{\emph{Proceedings of the {Eleventh} {ACM} {Conference} on
  {Recommender} {Systems}}} \emph{(\bibinfo{series}{{RecSys} '17})}.
  \bibinfo{publisher}{Association for Computing Machinery},
  \bibinfo{address}{New York, NY, USA}, \bibinfo{pages}{2}.
\newblock
\showISBNx{978-1-4503-4652-8}
\urldef\tempurl%
\url{https://doi.org/10.1145/3109859.3109870}
\showDOI{\tempurl}


\bibitem[Kaur et~al\mbox{.}(2019)]%
        {kaur_causal_2019}
\bibfield{author}{\bibinfo{person}{Prableen Kaur}, \bibinfo{person}{Agoritsa
  Polyzou}, {and} \bibinfo{person}{George Karypis}.}
  \bibinfo{year}{2019}\natexlab{}.
\newblock \showarticletitle{Causal {Inference} in {Higher} {Education}:
  {Building} {Better} {Curriculums}}. In \bibinfo{booktitle}{\emph{Proceedings
  of the {Sixth} (2019) {ACM} {Conference} on {Learning} @ {Scale}}}.
  \bibinfo{publisher}{ACM}, \bibinfo{address}{Chicago IL USA},
  \bibinfo{pages}{1--4}.
\newblock
\showISBNx{978-1-4503-6804-9}
\urldef\tempurl%
\url{https://doi.org/10.1145/3330430.3333663}
\showDOI{\tempurl}


\bibitem[Kim et~al\mbox{.}(2021)]%
        {kim_human-centered_2021}
\bibfield{author}{\bibinfo{person}{Seunghyun Kim}, \bibinfo{person}{Afsaneh
  Razi}, \bibinfo{person}{Gianluca Stringhini}, \bibinfo{person}{Pamela~J.
  Wisniewski}, {and} \bibinfo{person}{Munmun De~Choudhury}.}
  \bibinfo{year}{2021}\natexlab{}.
\newblock \showarticletitle{A {Human}-{Centered} {Systematic} {Literature}
  {Review} of {Cyberbullying} {Detection} {Algorithms}}.
\newblock \bibinfo{journal}{\emph{Proceedings of the ACM on Human-Computer
  Interaction}} \bibinfo{volume}{5}, \bibinfo{number}{CSCW2}
  (\bibinfo{date}{Oct.} \bibinfo{year}{2021}), \bibinfo{pages}{1--34}.
\newblock
\showISSN{2573-0142}
\urldef\tempurl%
\url{https://doi.org/10.1145/3476066}
\showDOI{\tempurl}


\bibitem[Kostopoulos et~al\mbox{.}(2015)]%
        {kostopoulos_estimating_2015}
\bibfield{author}{\bibinfo{person}{Georgios Kostopoulos},
  \bibinfo{person}{Sotiris Kotsiantis}, {and} \bibinfo{person}{Panagiotis
  Pintelas}.} \bibinfo{year}{2015}\natexlab{}.
\newblock \showarticletitle{Estimating student dropout in distance higher
  education using semi-supervised techniques}. In
  \bibinfo{booktitle}{\emph{Proceedings of the 19th {Panhellenic} {Conference}
  on {Informatics}}}. \bibinfo{publisher}{ACM}, \bibinfo{address}{Athens
  Greece}, \bibinfo{pages}{38--43}.
\newblock
\showISBNx{978-1-4503-3551-5}
\urldef\tempurl%
\url{https://doi.org/10.1145/2801948.2802013}
\showDOI{\tempurl}


\bibitem[Kumar et~al\mbox{.}(2017)]%
        {kumar_literature_2017}
\bibfield{author}{\bibinfo{person}{Mukesh Kumar}, \bibinfo{person}{A.J. Singh},
  {and} \bibinfo{person}{Disha Handa}.} \bibinfo{year}{2017}\natexlab{}.
\newblock \showarticletitle{Literature {Survey} on {Student}’s {Performance}
  {Prediction} in {Education} using {Data} {Mining} {Techniques}}.
\newblock \bibinfo{journal}{\emph{International Journal of Education and
  Management Engineering}} \bibinfo{volume}{7}, \bibinfo{number}{6}
  (\bibinfo{date}{Nov.} \bibinfo{year}{2017}), \bibinfo{pages}{40--49}.
\newblock
\showISSN{23053623, 23058463}
\urldef\tempurl%
\url{https://doi.org/10.5815/ijeme.2017.06.05}
\showDOI{\tempurl}


\bibitem[Kung and Yu(2020)]%
        {kung_interpretable_2020}
\bibfield{author}{\bibinfo{person}{Catherine Kung} {and}
  \bibinfo{person}{Renzhe Yu}.} \bibinfo{year}{2020}\natexlab{}.
\newblock \showarticletitle{Interpretable {Models} {Do} {Not} {Compromise}
  {Accuracy} or {Fairness} in {Predicting} {College} {Success}}. In
  \bibinfo{booktitle}{\emph{Proceedings of the {Seventh} {ACM} {Conference} on
  {Learning} @ {Scale}}}. \bibinfo{publisher}{ACM}, \bibinfo{address}{Virtual
  Event USA}, \bibinfo{pages}{413--416}.
\newblock
\showISBNx{978-1-4503-7951-9}
\urldef\tempurl%
\url{https://doi.org/10.1145/3386527.3406755}
\showDOI{\tempurl}


\bibitem[Kusner et~al\mbox{.}(2017)]%
        {kusner_counterfactual_2017}
\bibfield{author}{\bibinfo{person}{Matt~J Kusner}, \bibinfo{person}{Joshua
  Loftus}, \bibinfo{person}{Chris Russell}, {and} \bibinfo{person}{Ricardo
  Silva}.} \bibinfo{year}{2017}\natexlab{}.
\newblock \showarticletitle{Counterfactual {Fairness}}. In
  \bibinfo{booktitle}{\emph{Advances in {Neural} {Information} {Processing}
  {Systems}}}, Vol.~\bibinfo{volume}{30}. \bibinfo{publisher}{Curran
  Associates, Inc.}
\newblock
\urldef\tempurl%
\url{https://proceedings.neurips.cc/paper/2017/hash/a486cd07e4ac3d270571622f4f316ec5-Abstract.html}
\showURL{%
\tempurl}


\bibitem[Lang et~al\mbox{.}(2021)]%
        {lang_forecasting_2021}
\bibfield{author}{\bibinfo{person}{David Lang}, \bibinfo{person}{Alex Wang},
  \bibinfo{person}{Nathan Dalal}, \bibinfo{person}{Andreas Paepcke}, {and}
  \bibinfo{person}{Mitchell Stevens}.} \bibinfo{year}{2021}\natexlab{}.
\newblock \showarticletitle{Forecasting {Undergraduate} {Majors} {Using}
  {Academic} {Transcript} {Data}}. In \bibinfo{booktitle}{\emph{Proceedings of
  the {Eighth} {ACM} {Conference} on {Learning} @ {Scale}}}.
  \bibinfo{publisher}{ACM}, \bibinfo{address}{Virtual Event Germany},
  \bibinfo{pages}{243--246}.
\newblock
\showISBNx{978-1-4503-8215-1}
\urldef\tempurl%
\url{https://doi.org/10.1145/3430895.3460149}
\showDOI{\tempurl}


\bibitem[Lee and Kizilcec(2020)]%
        {lee_evaluation_2020}
\bibfield{author}{\bibinfo{person}{Hansol Lee} {and} \bibinfo{person}{René~F.
  Kizilcec}.} \bibinfo{year}{2020}\natexlab{}.
\newblock \bibinfo{booktitle}{\emph{Evaluation of {Fairness} {Trade}-offs in
  {Predicting} {Student} {Success}}}.
\newblock \bibinfo{type}{{T}echnical {R}eport} arXiv:2007.00088.
  \bibinfo{institution}{arXiv}.
\newblock
\urldef\tempurl%
\url{https://doi.org/10.48550/arXiv.2007.00088}
\showDOI{\tempurl}
\newblock
\shownote{arXiv:2007.00088 [cs] type: article}.


\bibitem[Li et~al\mbox{.}(2021)]%
        {li_yet_2021}
\bibfield{author}{\bibinfo{person}{Chenglu Li}, \bibinfo{person}{Wanli Xing},
  {and} \bibinfo{person}{Walter Leite}.} \bibinfo{year}{2021}\natexlab{}.
\newblock \showarticletitle{Yet {Another} {Predictive} {Model}? {Fair}
  {Predictions} of {Students}’ {Learning} {Outcomes} in an {Online} {Math}
  {Learning} {Platform}}. In \bibinfo{booktitle}{\emph{{LAK21}: 11th
  {International} {Learning} {Analytics} and {Knowledge} {Conference}}}.
  \bibinfo{publisher}{ACM}, \bibinfo{address}{Irvine CA USA},
  \bibinfo{pages}{572--578}.
\newblock
\showISBNx{978-1-4503-8935-8}
\urldef\tempurl%
\url{https://doi.org/10.1145/3448139.3448200}
\showDOI{\tempurl}


\bibitem[Luzardo et~al\mbox{.}(2014)]%
        {luzardo_estimation_2014}
\bibfield{author}{\bibinfo{person}{Gonzalo Luzardo}, \bibinfo{person}{Bruno
  Guamán}, \bibinfo{person}{Katherine Chiluiza}, \bibinfo{person}{Jaime
  Castells}, {and} \bibinfo{person}{Xavier Ochoa}.}
  \bibinfo{year}{2014}\natexlab{}.
\newblock \showarticletitle{Estimation of {Presentations} {Skills} {Based} on
  {Slides} and {Audio} {Features}}. In \bibinfo{booktitle}{\emph{Proceedings of
  the 2014 {ACM} workshop on {Multimodal} {Learning} {Analytics} {Workshop} and
  {Grand} {Challenge}}}. \bibinfo{publisher}{ACM}, \bibinfo{address}{Istanbul
  Turkey}, \bibinfo{pages}{37--44}.
\newblock
\showISBNx{978-1-4503-0488-7}
\urldef\tempurl%
\url{https://doi.org/10.1145/2666633.2666639}
\showDOI{\tempurl}


\bibitem[Ma et~al\mbox{.}(2018)]%
        {ma_using_2018}
\bibfield{author}{\bibinfo{person}{Xiaofeng Ma}, \bibinfo{person}{Yan Yang},
  {and} \bibinfo{person}{Zhurong Zhou}.} \bibinfo{year}{2018}\natexlab{}.
\newblock \showarticletitle{Using {Machine} {Learning} {Algorithm} to {Predict}
  {Student} {Pass} {Rates} {In} {Online} {Education}}. In
  \bibinfo{booktitle}{\emph{Proceedings of the 3rd {International} {Conference}
  on {Multimedia} {Systems} and {Signal} {Processing} - {ICMSSP} '18}}.
  \bibinfo{publisher}{ACM Press}, \bibinfo{address}{Shenzhen, China},
  \bibinfo{pages}{156--161}.
\newblock
\showISBNx{978-1-4503-6457-7}
\urldef\tempurl%
\url{https://doi.org/10.1145/3220162.3220188}
\showDOI{\tempurl}


\bibitem[Maramag and Palaoag(2019)]%
        {maramag_assessing_2019}
\bibfield{author}{\bibinfo{person}{Charlot~L. Maramag} {and}
  \bibinfo{person}{Thelma~D. Palaoag}.} \bibinfo{year}{2019}\natexlab{}.
\newblock \showarticletitle{Assessing {CSU} {Students}' {Academic}
  {Performance} on {iLearn} {Portal} {Using} {Data} {Analytics}}. In
  \bibinfo{booktitle}{\emph{Proceedings of the 2019 5th {International}
  {Conference} on {Computing} and {Artificial} {Intelligence} - {ICCAI} '19}}.
  \bibinfo{publisher}{ACM Press}, \bibinfo{address}{Bali, Indonesia},
  \bibinfo{pages}{25--29}.
\newblock
\showISBNx{978-1-4503-6106-4}
\urldef\tempurl%
\url{https://doi.org/10.1145/3330482.3330495}
\showDOI{\tempurl}


\bibitem[Marcinkowski et~al\mbox{.}(2020)]%
        {marcinkowski_implications_2020}
\bibfield{author}{\bibinfo{person}{Frank Marcinkowski}, \bibinfo{person}{Kimon
  Kieslich}, \bibinfo{person}{Christopher Starke}, {and} \bibinfo{person}{Marco
  Lünich}.} \bibinfo{year}{2020}\natexlab{}.
\newblock \showarticletitle{Implications of {AI} (un-)fairness in higher
  education admissions: the effects of perceived {AI} (un-)fairness on exit,
  voice and organizational reputation}. In
  \bibinfo{booktitle}{\emph{Proceedings of the 2020 {Conference} on {Fairness},
  {Accountability}, and {Transparency}}} \emph{(\bibinfo{series}{{FAT}* '20})}.
  \bibinfo{publisher}{Association for Computing Machinery},
  \bibinfo{address}{New York, NY, USA}, \bibinfo{pages}{122--130}.
\newblock
\showISBNx{978-1-4503-6936-7}
\urldef\tempurl%
\url{https://doi.org/10.1145/3351095.3372867}
\showDOI{\tempurl}


\bibitem[McPherson et~al\mbox{.}(2016)]%
        {mcpherson_student_2016}
\bibfield{author}{\bibinfo{person}{Jen McPherson}, \bibinfo{person}{Huong~Ly
  Tong}, \bibinfo{person}{Scott~J. Fatt}, {and} \bibinfo{person}{Danny Y.~T.
  Liu}.} \bibinfo{year}{2016}\natexlab{}.
\newblock \showarticletitle{Student perspectives on data provision and use:
  starting to unpack disciplinary differences}. In
  \bibinfo{booktitle}{\emph{Proceedings of the {Sixth} {International}
  {Conference} on {Learning} {Analytics} \& {Knowledge}}}
  \emph{(\bibinfo{series}{{LAK} '16})}. \bibinfo{publisher}{Association for
  Computing Machinery}, \bibinfo{address}{New York, NY, USA},
  \bibinfo{pages}{158--167}.
\newblock
\showISBNx{978-1-4503-4190-5}
\urldef\tempurl%
\url{https://doi.org/10.1145/2883851.2883945}
\showDOI{\tempurl}


\bibitem[Moher et~al\mbox{.}(2009)]%
        {moher_preferred_2009}
\bibfield{author}{\bibinfo{person}{David Moher}, \bibinfo{person}{Alessandro
  Liberati}, \bibinfo{person}{Jennifer Tetzlaff}, {and}
  \bibinfo{person}{Douglas~G. Altman}.} \bibinfo{year}{2009}\natexlab{}.
\newblock \showarticletitle{Preferred reporting items for systematic reviews
  and meta-analyses: the {PRISMA} statement}.
\newblock \bibinfo{journal}{\emph{BMJ}}  \bibinfo{volume}{339}
  (\bibinfo{date}{July} \bibinfo{year}{2009}), \bibinfo{pages}{b2535}.
\newblock
\showISSN{1756-1833}
\urldef\tempurl%
\url{https://doi.org/10.1136/bmj.b2535}
\showDOI{\tempurl}
\newblock
\shownote{Publisher: British Medical Journal Publishing Group Section: Research
  Methods \&amp; Reporting}.


\bibitem[Mouw and Khanna(1993)]%
        {mouw_prediction_1993}
\bibfield{author}{\bibinfo{person}{John~T. Mouw} {and} \bibinfo{person}{Ritu~K.
  Khanna}.} \bibinfo{year}{1993}\natexlab{}.
\newblock \showarticletitle{Prediction of academic success: {A} review of the
  literature and some recommendations}.
\newblock \bibinfo{journal}{\emph{College Student Journal}}
  \bibinfo{volume}{27}, \bibinfo{number}{3} (\bibinfo{year}{1993}),
  \bibinfo{pages}{328--336}.
\newblock
\showISSN{2691-3887}
\newblock
\shownote{Place: US Publisher: Project Innovation of Mobile}.


\bibitem[Muslim et~al\mbox{.}(2016)]%
        {muslim_rule-based_2016}
\bibfield{author}{\bibinfo{person}{Arham Muslim},
  \bibinfo{person}{Mohamed~Amine Chatti}, \bibinfo{person}{Tanmaya Mahapatra},
  {and} \bibinfo{person}{Ulrik Schroeder}.} \bibinfo{year}{2016}\natexlab{}.
\newblock \showarticletitle{A rule-based indicator definition tool for
  personalized learning analytics}. In \bibinfo{booktitle}{\emph{Proceedings of
  the {Sixth} {International} {Conference} on {Learning} {Analytics} \&
  {Knowledge} - {LAK} '16}}. \bibinfo{publisher}{ACM Press},
  \bibinfo{address}{Edinburgh, United Kingdom}, \bibinfo{pages}{264--273}.
\newblock
\showISBNx{978-1-4503-4190-5}
\urldef\tempurl%
\url{https://doi.org/10.1145/2883851.2883921}
\showDOI{\tempurl}


\bibitem[Nespereira et~al\mbox{.}(2014)]%
        {nespereira_is_2014}
\bibfield{author}{\bibinfo{person}{Celia~González Nespereira},
  \bibinfo{person}{Kais Dai}, \bibinfo{person}{Rebeca P.~Díaz Redondo}, {and}
  \bibinfo{person}{Ana~Fernández Vilas}.} \bibinfo{year}{2014}\natexlab{}.
\newblock \showarticletitle{Is the {LMS} access frequency a sign of students'
  success in face-to-face higher education?}. In
  \bibinfo{booktitle}{\emph{Proceedings of the {Second} {International}
  {Conference} on {Technological} {Ecosystems} for {Enhancing}
  {Multiculturality} - {TEEM} '14}}. \bibinfo{publisher}{ACM Press},
  \bibinfo{address}{Salamanca, Spain}, \bibinfo{pages}{283--290}.
\newblock
\showISBNx{978-1-4503-2896-8}
\urldef\tempurl%
\url{https://doi.org/10.1145/2669711.2669912}
\showDOI{\tempurl}


\bibitem[Obeid et~al\mbox{.}(2018)]%
        {obeid_ontology-based_2018}
\bibfield{author}{\bibinfo{person}{Charbel Obeid}, \bibinfo{person}{Inaya
  Lahoud}, \bibinfo{person}{Hicham El~Khoury}, {and}
  \bibinfo{person}{Pierre-Antoine Champin}.} \bibinfo{year}{2018}\natexlab{}.
\newblock \showarticletitle{Ontology-based {Recommender} {System} in {Higher}
  {Education}}. In \bibinfo{booktitle}{\emph{Companion of the {The} {Web}
  {Conference} 2018 on {The} {Web} {Conference} 2018 - {WWW} '18}}.
  \bibinfo{publisher}{ACM Press}, \bibinfo{address}{Lyon, France},
  \bibinfo{pages}{1031--1034}.
\newblock
\showISBNx{978-1-4503-5640-4}
\urldef\tempurl%
\url{https://doi.org/10.1145/3184558.3191533}
\showDOI{\tempurl}


\bibitem[Ofori et~al\mbox{.}(2020)]%
        {ofori_using_2020}
\bibfield{author}{\bibinfo{person}{Francis Ofori}, \bibinfo{person}{Elizaphan
  Maina}, {and} \bibinfo{person}{Rhoda Gitonga}.}
  \bibinfo{year}{2020}\natexlab{}.
\newblock \showarticletitle{Using {Machine} {Learning} {Algorithms} to
  {Predict} {Studentsâ}€™ {Performance} and {Improve} {Learning}
  {Outcome}: {A} {Literature} {Based} {Review}}.
\newblock \bibinfo{journal}{\emph{Journal of Information and Technology}}
  \bibinfo{volume}{4}, \bibinfo{number}{1} (\bibinfo{date}{March}
  \bibinfo{year}{2020}).
\newblock
\urldef\tempurl%
\url{https://stratfordjournals.org/journals/index.php/Journal-of-Information-and-Techn/article/view/480}
\showURL{%
\tempurl}
\newblock
\shownote{Number: 1}.


\bibitem[Ojajuni et~al\mbox{.}(2021)]%
        {ojajuni_predicting_2021}
\bibfield{author}{\bibinfo{person}{Opeyemi Ojajuni}, \bibinfo{person}{Foluso
  Ayeni}, \bibinfo{person}{Olagunju Akodu}, \bibinfo{person}{Femi Ekanoye},
  \bibinfo{person}{Samson Adewole}, \bibinfo{person}{Timothy Ayo},
  \bibinfo{person}{Sanjay Misra}, {and} \bibinfo{person}{Victor Mbarika}.}
  \bibinfo{year}{2021}\natexlab{}.
\newblock \showarticletitle{Predicting {Student} {Academic} {Performance}
  {Using} {Machine} {Learning}}. In \bibinfo{booktitle}{\emph{Computational
  {Science} and {Its} {Applications} – {ICCSA} 2021}}
  \emph{(\bibinfo{series}{Lecture {Notes} in {Computer} {Science}})},
  \bibfield{editor}{\bibinfo{person}{Osvaldo Gervasi},
  \bibinfo{person}{Beniamino Murgante}, \bibinfo{person}{Sanjay Misra},
  \bibinfo{person}{Chiara Garau}, \bibinfo{person}{Ivan Blečić},
  \bibinfo{person}{David Taniar}, \bibinfo{person}{Bernady~O. Apduhan},
  \bibinfo{person}{Ana Maria A.~C. Rocha}, \bibinfo{person}{Eufemia Tarantino},
  {and} \bibinfo{person}{Carmelo~Maria Torre}} (Eds.).
  \bibinfo{publisher}{Springer International Publishing},
  \bibinfo{address}{Cham}, \bibinfo{pages}{481--491}.
\newblock
\showISBNx{978-3-030-87013-3}
\urldef\tempurl%
\url{https://doi.org/10.1007/978-3-030-87013-3_36}
\showDOI{\tempurl}


\bibitem[Olssen~* and Peters(2005)]%
        {olssen__neoliberalism_2005}
\bibfield{author}{\bibinfo{person}{Mark Olssen~*} {and}
  \bibinfo{person}{Michael~A. Peters}.} \bibinfo{year}{2005}\natexlab{}.
\newblock \showarticletitle{Neoliberalism, higher education and the knowledge
  economy: from the free market to knowledge capitalism}.
\newblock \bibinfo{journal}{\emph{Journal of Education Policy}}
  \bibinfo{volume}{20}, \bibinfo{number}{3} (\bibinfo{date}{Jan.}
  \bibinfo{year}{2005}), \bibinfo{pages}{313--345}.
\newblock
\showISSN{0268-0939, 1464-5106}
\urldef\tempurl%
\url{https://doi.org/10.1080/02680930500108718}
\showDOI{\tempurl}


\bibitem[Oreshin et~al\mbox{.}(2020)]%
        {oreshin_implementing_2020}
\bibfield{author}{\bibinfo{person}{Svyatoslav Oreshin}, \bibinfo{person}{Andrey
  Filchenkov}, \bibinfo{person}{Polina Petrusha}, \bibinfo{person}{Egor
  Krasheninnikov}, \bibinfo{person}{Alexander Panfilov}, \bibinfo{person}{Igor
  Glukhov}, \bibinfo{person}{Yulia Kaliberda}, \bibinfo{person}{Daniil
  Masalskiy}, \bibinfo{person}{Alexey Serdyukov}, \bibinfo{person}{Vladimir
  Kazakovtsev}, \bibinfo{person}{Maksim Khlopotov}, \bibinfo{person}{Timofey
  Podolenchuk}, \bibinfo{person}{Ivan Smetannikov}, {and}
  \bibinfo{person}{Daria Kozlova}.} \bibinfo{year}{2020}\natexlab{}.
\newblock \showarticletitle{Implementing a {Machine} {Learning} {Approach} to
  {Predicting} {Students}’ {Academic} {Outcomes}}. In
  \bibinfo{booktitle}{\emph{2020 {International} {Conference} on {Control},
  {Robotics} and {Intelligent} {System}}}. \bibinfo{publisher}{ACM},
  \bibinfo{address}{Xiamen China}, \bibinfo{pages}{78--83}.
\newblock
\showISBNx{978-1-4503-8805-4}
\urldef\tempurl%
\url{https://doi.org/10.1145/3437802.3437816}
\showDOI{\tempurl}


\bibitem[Orong et~al\mbox{.}(2020)]%
        {orong_predictive_2020}
\bibfield{author}{\bibinfo{person}{Markdy~Y. Orong},
  \bibinfo{person}{Roseclaremath~A. Caroro}, \bibinfo{person}{Geraldine~D.
  Durias}, \bibinfo{person}{Joey~A. Cabrera}, \bibinfo{person}{Herwina Lonzon},
  {and} \bibinfo{person}{Gretel~T. Ricalde}.} \bibinfo{year}{2020}\natexlab{}.
\newblock \showarticletitle{A {Predictive} {Analytics} {Approach} in
  {Determining} the {Predictors} of {Student} {Attrition} in the {Higher}
  {Education} {Institutions} in the {Philippines}}. In
  \bibinfo{booktitle}{\emph{Proceedings of the 3rd {International} {Conference}
  on {Software} {Engineering} and {Information} {Management}}}.
  \bibinfo{publisher}{ACM}, \bibinfo{address}{Sydney NSW Australia},
  \bibinfo{pages}{222--225}.
\newblock
\showISBNx{978-1-4503-7690-7}
\urldef\tempurl%
\url{https://doi.org/10.1145/3378936.3378956}
\showDOI{\tempurl}


\bibitem[Otoo-Arthur and Van~Zyl(2019)]%
        {otoo-arthur_systematic_2019}
\bibfield{author}{\bibinfo{person}{David Otoo-Arthur} {and}
  \bibinfo{person}{Terence Van~Zyl}.} \bibinfo{year}{2019}\natexlab{}.
\newblock \showarticletitle{A {Systematic} {Review} on {Big} {Data} {Analytics}
  {Frameworks} for {Higher} {Education} - {Tools} and {Algorithms}}. In
  \bibinfo{booktitle}{\emph{Proceedings of the 2019 2nd {International}
  {Conference} on {E}-{Business}, {Information} {Management} and {Computer}
  {Science}}}. \bibinfo{publisher}{ACM}, \bibinfo{address}{Kuala Lumpur
  Malaysia}, \bibinfo{pages}{1--9}.
\newblock
\showISBNx{978-1-4503-6649-6}
\urldef\tempurl%
\url{https://doi.org/10.1145/3377817.3377836}
\showDOI{\tempurl}


\bibitem[Pardos et~al\mbox{.}(2019)]%
        {pardos_data-assistive_2019}
\bibfield{author}{\bibinfo{person}{Zachary~A. Pardos}, \bibinfo{person}{Hung
  Chau}, {and} \bibinfo{person}{Haocheng Zhao}.}
  \bibinfo{year}{2019}\natexlab{}.
\newblock \showarticletitle{Data-{Assistive} {Course}-to-{Course}
  {Articulation} {Using} {Machine} {Translation}}. In
  \bibinfo{booktitle}{\emph{Proceedings of the {Sixth} (2019) {ACM}
  {Conference} on {Learning} @ {Scale}}}. \bibinfo{publisher}{ACM},
  \bibinfo{address}{Chicago IL USA}, \bibinfo{pages}{1--10}.
\newblock
\showISBNx{978-1-4503-6804-9}
\urldef\tempurl%
\url{https://doi.org/10.1145/3330430.3333622}
\showDOI{\tempurl}


\bibitem[Pardos and Jiang(2020)]%
        {pardos_designing_2020}
\bibfield{author}{\bibinfo{person}{Zachary~A. Pardos} {and}
  \bibinfo{person}{Weijie Jiang}.} \bibinfo{year}{2020}\natexlab{}.
\newblock \showarticletitle{Designing for serendipity in a university course
  recommendation system}. In \bibinfo{booktitle}{\emph{Proceedings of the
  {Tenth} {International} {Conference} on {Learning} {Analytics} \&
  {Knowledge}}}. \bibinfo{publisher}{ACM}, \bibinfo{address}{Frankfurt
  Germany}, \bibinfo{pages}{350--359}.
\newblock
\showISBNx{978-1-4503-7712-6}
\urldef\tempurl%
\url{https://doi.org/10.1145/3375462.3375524}
\showDOI{\tempurl}


\bibitem[Perez-Sanagustin et~al\mbox{.}(2021)]%
        {perez-sanagustin_can_2021}
\bibfield{author}{\bibinfo{person}{Mar Perez-Sanagustin},
  \bibinfo{person}{Ronald Pérez-Álvarez}, \bibinfo{person}{Jorge
  Maldonado-Mahauad}, \bibinfo{person}{Esteban Villalobos},
  \bibinfo{person}{Isabel Hilliger}, \bibinfo{person}{Josefina Hernández},
  \bibinfo{person}{Diego Sapunar}, \bibinfo{person}{Pedro~Manuel
  Moreno-Marcos}, \bibinfo{person}{Pedro~J. Muñoz-Merino},
  \bibinfo{person}{Carlos Delgado~Kloos}, {and} \bibinfo{person}{Jon Imaz}.}
  \bibinfo{year}{2021}\natexlab{}.
\newblock \showarticletitle{Can {Feedback} based on {Predictive} {Data}
  {Improve} {Learners}' {Passing} {Rates} in {MOOCs}? {A} {Preliminary}
  {Analysis}}. In \bibinfo{booktitle}{\emph{Proceedings of the {Eighth} {ACM}
  {Conference} on {Learning} @ {Scale}}}. \bibinfo{publisher}{ACM},
  \bibinfo{address}{Virtual Event Germany}, \bibinfo{pages}{339--342}.
\newblock
\showISBNx{978-1-4503-8215-1}
\urldef\tempurl%
\url{https://doi.org/10.1145/3430895.3460991}
\showDOI{\tempurl}


\bibitem[Pokorny(2021)]%
        {pokorny_machine_2021}
\bibfield{author}{\bibinfo{person}{Kian~L. Pokorny}.}
  \bibinfo{year}{2021}\natexlab{}.
\newblock \showarticletitle{A machine learning approach to understanding the
  viability of private 4-year higher-education institutions}.
\newblock \bibinfo{journal}{\emph{Journal of Computing Sciences in Colleges}}
  \bibinfo{volume}{37}, \bibinfo{number}{4} (\bibinfo{date}{Oct.}
  \bibinfo{year}{2021}), \bibinfo{pages}{50--57}.
\newblock
\showISSN{1937-4771}


\bibitem[Prinsloo and Slade(2017)]%
        {prinsloo_elephant_2017}
\bibfield{author}{\bibinfo{person}{Paul Prinsloo} {and} \bibinfo{person}{Sharon
  Slade}.} \bibinfo{year}{2017}\natexlab{}.
\newblock \showarticletitle{An elephant in the learning analytics room: the
  obligation to act}. In \bibinfo{booktitle}{\emph{Proceedings of the {Seventh}
  {International} {Learning} {Analytics} \& {Knowledge} {Conference}}}
  \emph{(\bibinfo{series}{{LAK} '17})}. \bibinfo{publisher}{Association for
  Computing Machinery}, \bibinfo{address}{New York, NY, USA},
  \bibinfo{pages}{46--55}.
\newblock
\showISBNx{978-1-4503-4870-6}
\urldef\tempurl%
\url{https://doi.org/10.1145/3027385.3027406}
\showDOI{\tempurl}


\bibitem[Razi et~al\mbox{.}(2021)]%
        {razi_human-centered_2021}
\bibfield{author}{\bibinfo{person}{Afsaneh Razi}, \bibinfo{person}{Seunghyun
  Kim}, \bibinfo{person}{Ashwaq Alsoubai}, \bibinfo{person}{Gianluca
  Stringhini}, \bibinfo{person}{Thamar Solorio}, \bibinfo{person}{Munmun
  De~Choudhury}, {and} \bibinfo{person}{Pamela~J. Wisniewski}.}
  \bibinfo{year}{2021}\natexlab{}.
\newblock \showarticletitle{A {Human}-{Centered} {Systematic} {Literature}
  {Review} of the {Computational} {Approaches} for {Online} {Sexual} {Risk}
  {Detection}}.
\newblock \bibinfo{journal}{\emph{Proceedings of the ACM on Human-Computer
  Interaction}} \bibinfo{volume}{5}, \bibinfo{number}{CSCW2}
  (\bibinfo{date}{Oct.} \bibinfo{year}{2021}), \bibinfo{pages}{465:1--465:38}.
\newblock
\urldef\tempurl%
\url{https://doi.org/10.1145/3479609}
\showDOI{\tempurl}


\bibitem[Rijati et~al\mbox{.}(2018)]%
        {rijati_multi-attribute_2018}
\bibfield{author}{\bibinfo{person}{Nova Rijati}, \bibinfo{person}{Surya
  Sumpeno}, {and} \bibinfo{person}{Mauridhi~Hery Purnomo}.}
  \bibinfo{year}{2018}\natexlab{}.
\newblock \showarticletitle{Multi-{Attribute} {Clustering} of {Student}'s
  {Entrepreneurial} {Potential} {Mapping} {Based} on {Its} {Characteristics}
  and the {Affecting} {Factors}: {Preliminary} {Study} on {Indonesian} {Higher}
  {Education} {Database}}. In \bibinfo{booktitle}{\emph{Proceedings of the 2018
  10th {International} {Conference} on {Computer} and {Automation}
  {Engineering}}}. \bibinfo{publisher}{ACM}, \bibinfo{address}{Brisbane
  Australia}, \bibinfo{pages}{11--16}.
\newblock
\showISBNx{978-1-4503-6410-2}
\urldef\tempurl%
\url{https://doi.org/10.1145/3192975.3193014}
\showDOI{\tempurl}


\bibitem[Rogers et~al\mbox{.}(2014)]%
        {rogers_modest_2014}
\bibfield{author}{\bibinfo{person}{Tim Rogers}, \bibinfo{person}{Cassandra
  Colvin}, {and} \bibinfo{person}{Belinda Chiera}.}
  \bibinfo{year}{2014}\natexlab{}.
\newblock \showarticletitle{Modest analytics: using the index method to
  identify students at risk of failure}. In
  \bibinfo{booktitle}{\emph{Proceedings of the {Fourth} {International}
  {Conference} on {Learning} {Analytics} {And} {Knowledge}}}.
  \bibinfo{publisher}{ACM}, \bibinfo{address}{Indianapolis Indiana USA},
  \bibinfo{pages}{118--122}.
\newblock
\showISBNx{978-1-4503-2664-3}
\urldef\tempurl%
\url{https://doi.org/10.1145/2567574.2567629}
\showDOI{\tempurl}


\bibitem[Rudin(2019)]%
        {rudin_stop_2019}
\bibfield{author}{\bibinfo{person}{Cynthia Rudin}.}
  \bibinfo{year}{2019}\natexlab{}.
\newblock \showarticletitle{Stop explaining black box machine learning models
  for high stakes decisions and use interpretable models instead}.
\newblock \bibinfo{journal}{\emph{Nature Machine Intelligence}}
  \bibinfo{volume}{1}, \bibinfo{number}{5} (\bibinfo{date}{May}
  \bibinfo{year}{2019}), \bibinfo{pages}{206--215}.
\newblock
\showISSN{2522-5839}
\urldef\tempurl%
\url{https://doi.org/10.1038/s42256-019-0048-x}
\showDOI{\tempurl}


\bibitem[Sallaberry et~al\mbox{.}(2021)]%
        {sallaberry_comparison_2021}
\bibfield{author}{\bibinfo{person}{Lucas~H. Sallaberry},
  \bibinfo{person}{Romero Tori}, {and} \bibinfo{person}{Fatima L~S Nunes}.}
  \bibinfo{year}{2021}\natexlab{}.
\newblock \showarticletitle{Comparison of machine learning algorithms for
  automatic assessment of performance in a virtual reality dental simulator}.
  In \bibinfo{booktitle}{\emph{Symposium on {Virtual} and {Augmented}
  {Reality}}}. \bibinfo{publisher}{ACM}, \bibinfo{address}{Virtual Event
  Brazil}, \bibinfo{pages}{14--23}.
\newblock
\showISBNx{978-1-4503-9552-6}
\urldef\tempurl%
\url{https://doi.org/10.1145/3488162.3488207}
\showDOI{\tempurl}


\bibitem[Sandra et~al\mbox{.}(2021)]%
        {sandra_machine_2021}
\bibfield{author}{\bibinfo{person}{Lidia Sandra}, \bibinfo{person}{Ford
  Lumbangaol}, {and} \bibinfo{person}{Tokuro Matsuo}.}
  \bibinfo{year}{2021}\natexlab{}.
\newblock \showarticletitle{Machine {Learning} {Algorithm} to {Predict}
  {Student}’s {Performance}: {A} {Systematic} {Literature} {Review}}.
\newblock \bibinfo{journal}{\emph{TEM Journal}} (\bibinfo{date}{Nov.}
  \bibinfo{year}{2021}), \bibinfo{pages}{1919--1927}.
\newblock
\showISSN{2217-8333, 2217-8309}
\urldef\tempurl%
\url{https://doi.org/10.18421/TEM104-56}
\showDOI{\tempurl}


\bibitem[Saxena et~al\mbox{.}(2020)]%
        {saxena_human-centered_2020}
\bibfield{author}{\bibinfo{person}{Devansh Saxena}, \bibinfo{person}{Karla
  Badillo-Urquiola}, \bibinfo{person}{Pamela~J. Wisniewski}, {and}
  \bibinfo{person}{Shion Guha}.} \bibinfo{year}{2020}\natexlab{}.
\newblock \showarticletitle{A {Human}-{Centered} {Review} of {Algorithms} used
  within the {U}.{S}. {Child} {Welfare} {System}}. In
  \bibinfo{booktitle}{\emph{Proceedings of the 2020 {CHI} {Conference} on
  {Human} {Factors} in {Computing} {Systems}}}. \bibinfo{publisher}{ACM},
  \bibinfo{address}{Honolulu HI USA}, \bibinfo{pages}{1--15}.
\newblock
\showISBNx{978-1-4503-6708-0}
\urldef\tempurl%
\url{https://doi.org/10.1145/3313831.3376229}
\showDOI{\tempurl}


\bibitem[Saxena and Guha(2020)]%
        {saxena_conducting_2020}
\bibfield{author}{\bibinfo{person}{Devansh Saxena} {and} \bibinfo{person}{Shion
  Guha}.} \bibinfo{year}{2020}\natexlab{}.
\newblock \showarticletitle{Conducting {Participatory} {Design} to {Improve}
  {Algorithms} in {Public} {Services}: {Lessons} and {Challenges}}. In
  \bibinfo{booktitle}{\emph{Conference {Companion} {Publication} of the 2020 on
  {Computer} {Supported} {Cooperative} {Work} and {Social} {Computing}}}
  \emph{(\bibinfo{series}{{CSCW} '20 {Companion}})}.
  \bibinfo{publisher}{Association for Computing Machinery},
  \bibinfo{address}{New York, NY, USA}, \bibinfo{pages}{383--388}.
\newblock
\showISBNx{978-1-4503-8059-1}
\urldef\tempurl%
\url{https://doi.org/10.1145/3406865.3418331}
\showDOI{\tempurl}


\bibitem[Sekeroglu et~al\mbox{.}(2019)]%
        {sekeroglu_student_2019}
\bibfield{author}{\bibinfo{person}{Boran Sekeroglu}, \bibinfo{person}{Kamil
  Dimililer}, {and} \bibinfo{person}{Kubra Tuncal}.}
  \bibinfo{year}{2019}\natexlab{}.
\newblock \showarticletitle{Student {Performance} {Prediction} and
  {Classification} {Using} {Machine} {Learning} {Algorithms}}. In
  \bibinfo{booktitle}{\emph{Proceedings of the 2019 8th {International}
  {Conference} on {Educational} and {Information} {Technology}}}.
  \bibinfo{publisher}{ACM}, \bibinfo{address}{Cambridge United Kingdom},
  \bibinfo{pages}{7--11}.
\newblock
\showISBNx{978-1-4503-6267-2}
\urldef\tempurl%
\url{https://doi.org/10.1145/3318396.3318419}
\showDOI{\tempurl}


\bibitem[Self(2014)]%
        {self_governance_2014}
\bibfield{author}{\bibinfo{person}{Richard~J. Self}.}
  \bibinfo{year}{2014}\natexlab{}.
\newblock \showarticletitle{Governance {Strategies} for the {Cloud}, {Big}
  {Data}, and {Other} {Technologies} in {Education}}. In
  \bibinfo{booktitle}{\emph{Proceedings of the 2014 {IEEE}/{ACM} 7th
  {International} {Conference} on {Utility} and {Cloud} {Computing}}}
  \emph{(\bibinfo{series}{{UCC} '14})}. \bibinfo{publisher}{IEEE Computer
  Society}, \bibinfo{address}{USA}, \bibinfo{pages}{630--635}.
\newblock
\showISBNx{978-1-4799-7881-6}
\urldef\tempurl%
\url{https://doi.org/10.1109/UCC.2014.101}
\showDOI{\tempurl}


\bibitem[Shahiri et~al\mbox{.}(2015)]%
        {shahiri_review_2015}
\bibfield{author}{\bibinfo{person}{Amirah~Mohamed Shahiri},
  \bibinfo{person}{Wahidah Husain}, {and} \bibinfo{person}{Nur’aini~Abdul
  Rashid}.} \bibinfo{year}{2015}\natexlab{}.
\newblock \showarticletitle{A {Review} on {Predicting} {Student}'s
  {Performance} {Using} {Data} {Mining} {Techniques}}.
\newblock \bibinfo{journal}{\emph{Procedia Computer Science}}
  \bibinfo{volume}{72} (\bibinfo{date}{Jan.} \bibinfo{year}{2015}),
  \bibinfo{pages}{414--422}.
\newblock
\showISSN{1877-0509}
\urldef\tempurl%
\url{https://doi.org/10.1016/j.procs.2015.12.157}
\showDOI{\tempurl}


\bibitem[Shneiderman(2021)]%
        {shneiderman_human-centered_2021}
\bibfield{author}{\bibinfo{person}{Ben Shneiderman}.}
  \bibinfo{year}{2021}\natexlab{}.
\newblock \showarticletitle{Human-{Centered} {AI}}.
\newblock \bibinfo{journal}{\emph{Issues in Science and Technology}}
  \bibinfo{volume}{37}, \bibinfo{number}{2} (\bibinfo{year}{2021}),
  \bibinfo{pages}{56--61}.
\newblock
\showISSN{07485492}
\urldef\tempurl%
\url{https://www.proquest.com/docview/2481224128?parentSessionId=chnCbaQ2NZxuTZjdR%2FSmni%2BibODmNifp1bKKqe%2FEjkA%3D&pq-origsite=primo&}
\showURL{%
\tempurl}
\newblock
\shownote{Num Pages: 56-61 Place: Washington, United States Publisher: Issues
  in Science and Technology}.


\bibitem[Shneiderman(2022)]%
        {shneiderman_human-centered_2022}
\bibfield{author}{\bibinfo{person}{Ben Shneiderman}.}
  \bibinfo{year}{2022}\natexlab{}.
\newblock \bibinfo{booktitle}{\emph{Human-{Centered} {AI}}}.
\newblock \bibinfo{publisher}{Oxford University Press},
  \bibinfo{address}{Oxford, New York}.
\newblock
\showISBNx{978-0-19-284529-0}


\bibitem[Slade and Prinsloo(2013)]%
        {slade_learning_2013}
\bibfield{author}{\bibinfo{person}{Sharon Slade} {and} \bibinfo{person}{Paul
  Prinsloo}.} \bibinfo{year}{2013}\natexlab{}.
\newblock \showarticletitle{Learning {Analytics}: {Ethical} {Issues} and
  {Dilemmas}}.
\newblock \bibinfo{journal}{\emph{American Behavioral Scientist}}
  \bibinfo{volume}{57}, \bibinfo{number}{10} (\bibinfo{date}{Oct.}
  \bibinfo{year}{2013}), \bibinfo{pages}{1510--1529}.
\newblock
\showISSN{0002-7642}
\urldef\tempurl%
\url{https://doi.org/10.1177/0002764213479366}
\showDOI{\tempurl}
\newblock
\shownote{Publisher: SAGE Publications Inc}.


\bibitem[Soleimani and Lee(2021)]%
        {soleimani_comparative_2021}
\bibfield{author}{\bibinfo{person}{Farahnaz Soleimani} {and}
  \bibinfo{person}{Jeonghyun Lee}.} \bibinfo{year}{2021}\natexlab{}.
\newblock \showarticletitle{Comparative {Analysis} of the {Feature}
  {Extraction} {Approaches} for {Predicting} {Learners} {Progress} in {Online}
  {Courses}: {MicroMasters} {Credential} versus {Traditional} {MOOCs}}. In
  \bibinfo{booktitle}{\emph{Proceedings of the {Eighth} {ACM} {Conference} on
  {Learning} @ {Scale}}} \emph{(\bibinfo{series}{L@{S} '21})}.
  \bibinfo{publisher}{Association for Computing Machinery},
  \bibinfo{address}{New York, NY, USA}, \bibinfo{pages}{151--159}.
\newblock
\showISBNx{978-1-4503-8215-1}
\urldef\tempurl%
\url{https://doi.org/10.1145/3430895.3460143}
\showDOI{\tempurl}


\bibitem[Staudaher et~al\mbox{.}(2020)]%
        {staudaher_predicting_2020}
\bibfield{author}{\bibinfo{person}{Shawn Staudaher}, \bibinfo{person}{Jeonghyun
  Lee}, {and} \bibinfo{person}{Farahnaz Soleimani}.}
  \bibinfo{year}{2020}\natexlab{}.
\newblock \showarticletitle{Predicting {Applicant} {Admission} {Status} for
  {Georgia} {Tech}'s {Online} {Master}'s in {Analytics} {Program}}. In
  \bibinfo{booktitle}{\emph{Proceedings of the {Seventh} {ACM} {Conference} on
  {Learning} @ {Scale}}}. \bibinfo{publisher}{ACM}, \bibinfo{address}{Virtual
  Event USA}, \bibinfo{pages}{309--312}.
\newblock
\showISBNx{978-1-4503-7951-9}
\urldef\tempurl%
\url{https://doi.org/10.1145/3386527.3406735}
\showDOI{\tempurl}


\bibitem[Stockton(2020)]%
        {stockton_incident_2020}
\bibfield{author}{\bibinfo{person}{Nick Stockton}.}
  \bibinfo{year}{2020}\natexlab{}.
\newblock \bibinfo{title}{Incident 78: {Meet} the {Secret} {Algorithm} {That}'s
  {Keeping} {Students} {Out} of {College}}.
\newblock
\newblock
\urldef\tempurl%
\url{https://incidentdatabase.ai/cite/78}
\showURL{%
\tempurl}


\bibitem[Trandafili et~al\mbox{.}(2012)]%
        {trandafili_discovery_2012}
\bibfield{author}{\bibinfo{person}{Evis Trandafili}, \bibinfo{person}{Alban
  Allkoçi}, \bibinfo{person}{Elinda Kajo}, {and} \bibinfo{person}{Aleksandër
  Xhuvani}.} \bibinfo{year}{2012}\natexlab{}.
\newblock \showarticletitle{Discovery and evaluation of student's profiles with
  machine learning}. In \bibinfo{booktitle}{\emph{Proceedings of the {Fifth}
  {Balkan} {Conference} in {Informatics} on - {BCI} '12}}.
  \bibinfo{publisher}{ACM Press}, \bibinfo{address}{Novi Sad, Serbia},
  \bibinfo{pages}{174}.
\newblock
\showISBNx{978-1-4503-1240-0}
\urldef\tempurl%
\url{https://doi.org/10.1145/2371316.2371350}
\showDOI{\tempurl}


\bibitem[Van~Goidsenhoven et~al\mbox{.}(2020)]%
        {van_goidsenhoven_predicting_2020}
\bibfield{author}{\bibinfo{person}{Steven Van~Goidsenhoven},
  \bibinfo{person}{Daria Bogdanova}, \bibinfo{person}{Galina Deeva},
  \bibinfo{person}{Seppe~vanden Broucke}, \bibinfo{person}{Jochen De~Weerdt},
  {and} \bibinfo{person}{Monique Snoeck}.} \bibinfo{year}{2020}\natexlab{}.
\newblock \showarticletitle{Predicting student success in a blended learning
  environment}. In \bibinfo{booktitle}{\emph{Proceedings of the {Tenth}
  {International} {Conference} on {Learning} {Analytics} \& {Knowledge}}}.
  \bibinfo{publisher}{ACM}, \bibinfo{address}{Frankfurt Germany},
  \bibinfo{pages}{17--25}.
\newblock
\showISBNx{978-1-4503-7712-6}
\urldef\tempurl%
\url{https://doi.org/10.1145/3375462.3375494}
\showDOI{\tempurl}


\bibitem[Vaughan and Wallach(2022)]%
        {vaughan_human-centered_2022}
\bibfield{author}{\bibinfo{person}{Jennifer~Wortman Vaughan} {and}
  \bibinfo{person}{Hanna Wallach}.} \bibinfo{year}{2022}\natexlab{}.
\newblock \showarticletitle{A {Human}-{Centered} {Agenda} for {Intelligible}
  {Machine} {Learning}}.
\newblock  (\bibinfo{date}{Sept.} \bibinfo{year}{2022}).
\newblock
\urldef\tempurl%
\url{https://www.microsoft.com/en-us/research/publication/a-human-centered-agenda-for-intelligible-machine-learning/}
\showURL{%
\tempurl}


\bibitem[Wang et~al\mbox{.}(2022)]%
        {wang_first-gen_2022}
\bibfield{author}{\bibinfo{person}{Weichen Wang}, \bibinfo{person}{Subigya
  Nepal}, \bibinfo{person}{Jeremy~F. Huckins}, \bibinfo{person}{Lessley
  Hernandez}, \bibinfo{person}{Vlado Vojdanovski}, \bibinfo{person}{Dante
  Mack}, \bibinfo{person}{Jane Plomp}, \bibinfo{person}{Arvind Pillai},
  \bibinfo{person}{Mikio Obuchi}, \bibinfo{person}{Alex daSilva},
  \bibinfo{person}{Eilis Murphy}, \bibinfo{person}{Elin Hedlund},
  \bibinfo{person}{Courtney Rogers}, \bibinfo{person}{Meghan Meyer}, {and}
  \bibinfo{person}{Andrew Campbell}.} \bibinfo{year}{2022}\natexlab{}.
\newblock \showarticletitle{First-{Gen} {Lens}: {Assessing} {Mental} {Health}
  of {First}-{Generation} {Students} across {Their} {First} {Year} at {College}
  {Using} {Mobile} {Sensing}}.
\newblock \bibinfo{journal}{\emph{Proceedings of the ACM on Interactive,
  Mobile, Wearable and Ubiquitous Technologies}} \bibinfo{volume}{6},
  \bibinfo{number}{2} (\bibinfo{date}{July} \bibinfo{year}{2022}),
  \bibinfo{pages}{1--32}.
\newblock
\showISSN{2474-9567}
\urldef\tempurl%
\url{https://doi.org/10.1145/3543194}
\showDOI{\tempurl}


\bibitem[Wham(2017)]%
        {wham_forecasting_2017}
\bibfield{author}{\bibinfo{person}{Drew Wham}.}
  \bibinfo{year}{2017}\natexlab{}.
\newblock \showarticletitle{Forecasting student outcomes at university-wide
  scale using machine learning}. In \bibinfo{booktitle}{\emph{Proceedings of
  the {Seventh} {International} {Learning} {Analytics} \& {Knowledge}
  {Conference}}}. \bibinfo{publisher}{ACM}, \bibinfo{address}{Vancouver British
  Columbia Canada}, \bibinfo{pages}{576--577}.
\newblock
\showISBNx{978-1-4503-4870-6}
\urldef\tempurl%
\url{https://doi.org/10.1145/3027385.3029467}
\showDOI{\tempurl}


\bibitem[Williamson and Kizilcec(2021)]%
        {williamson_learning_2021}
\bibfield{author}{\bibinfo{person}{Kimberly Williamson} {and}
  \bibinfo{person}{René~F. Kizilcec}.} \bibinfo{year}{2021}\natexlab{}.
\newblock \showarticletitle{Learning {Analytics} {Dashboard} {Research} {Has}
  {Neglected} {Diversity}, {Equity} and {Inclusion}}. In
  \bibinfo{booktitle}{\emph{Proceedings of the {Eighth} {ACM} {Conference} on
  {Learning} @ {Scale}}} \emph{(\bibinfo{series}{L@{S} '21})}.
  \bibinfo{publisher}{Association for Computing Machinery},
  \bibinfo{address}{New York, NY, USA}, \bibinfo{pages}{287--290}.
\newblock
\showISBNx{978-1-4503-8215-1}
\urldef\tempurl%
\url{https://doi.org/10.1145/3430895.3460160}
\showDOI{\tempurl}


\bibitem[Winters(1971)]%
        {winters_scheduling_1971}
\bibfield{author}{\bibinfo{person}{W.~K. Winters}.}
  \bibinfo{year}{1971}\natexlab{}.
\newblock \showarticletitle{A scheduling algorithm for a computer assisted
  registration system}.
\newblock \bibinfo{journal}{\emph{Commun. ACM}} \bibinfo{volume}{14},
  \bibinfo{number}{3} (\bibinfo{date}{March} \bibinfo{year}{1971}),
  \bibinfo{pages}{166--171}.
\newblock
\showISSN{0001-0782, 1557-7317}
\urldef\tempurl%
\url{https://doi.org/10.1145/362566.362569}
\showDOI{\tempurl}


\bibitem[Wobbrock and Kientz(2016)]%
        {wobbrock_research_2016}
\bibfield{author}{\bibinfo{person}{Jacob~O. Wobbrock} {and}
  \bibinfo{person}{Julie~A. Kientz}.} \bibinfo{year}{2016}\natexlab{}.
\newblock \showarticletitle{Research contributions in human-computer
  interaction}.
\newblock \bibinfo{journal}{\emph{Interactions}} \bibinfo{volume}{23},
  \bibinfo{number}{3} (\bibinfo{date}{April} \bibinfo{year}{2016}),
  \bibinfo{pages}{38--44}.
\newblock
\showISSN{1072-5520, 1558-3449}
\urldef\tempurl%
\url{https://doi.org/10.1145/2907069}
\showDOI{\tempurl}


\bibitem[Yan(2021)]%
        {yan_trends_2021}
\bibfield{author}{\bibinfo{person}{Hongli Yan}.}
  \bibinfo{year}{2021}\natexlab{}.
\newblock \showarticletitle{The {Trends} and {Challenges} of {Emerging}
  {Technologies} in {Higher} {Education}}. In \bibinfo{booktitle}{\emph{2021
  2nd {International} {Conference} on {Education} {Development} and
  {Studies}}}. \bibinfo{publisher}{ACM}, \bibinfo{address}{Hilo HI USA},
  \bibinfo{pages}{89--95}.
\newblock
\showISBNx{978-1-4503-8961-7}
\urldef\tempurl%
\url{https://doi.org/10.1145/3459043.3459060}
\showDOI{\tempurl}


\bibitem[Yu et~al\mbox{.}(2021)]%
        {yu_should_2021}
\bibfield{author}{\bibinfo{person}{Renzhe Yu}, \bibinfo{person}{Hansol Lee},
  {and} \bibinfo{person}{René~F. Kizilcec}.} \bibinfo{year}{2021}\natexlab{}.
\newblock \showarticletitle{Should {College} {Dropout} {Prediction} {Models}
  {Include} {Protected} {Attributes}?}. In
  \bibinfo{booktitle}{\emph{Proceedings of the {Eighth} {ACM} {Conference} on
  {Learning} @ {Scale}}}. \bibinfo{publisher}{ACM}, \bibinfo{address}{Virtual
  Event Germany}, \bibinfo{pages}{91--100}.
\newblock
\showISBNx{978-1-4503-8215-1}
\urldef\tempurl%
\url{https://doi.org/10.1145/3430895.3460139}
\showDOI{\tempurl}


\bibitem[Yu et~al\mbox{.}(2020)]%
        {yu_towards_2020}
\bibfield{author}{\bibinfo{person}{Renzhe Yu}, \bibinfo{person}{Qiujie Li},
  {and} \bibinfo{person}{Christian Fischer}.} \bibinfo{year}{2020}\natexlab{}.
\newblock \showarticletitle{Towards {Accurate} and {Fair} {Prediction} of
  {College} {Success}: {Evaluating} {Different} {Sources} of {Student} {Data}}.
  In \bibinfo{booktitle}{\emph{Educational {Data} {Mining}}}.
  \bibinfo{address}{Online}, \bibinfo{pages}{10}.
\newblock


\bibitem[Yu and Jo(2014)]%
        {yu_educational_2014}
\bibfield{author}{\bibinfo{person}{Taeho Yu} {and} \bibinfo{person}{Il-Hyun
  Jo}.} \bibinfo{year}{2014}\natexlab{}.
\newblock \showarticletitle{Educational technology approach toward learning
  analytics: relationship between student online behavior and learning
  performance in higher education}. In \bibinfo{booktitle}{\emph{Proceedings of
  the {Fourth} {International} {Conference} on {Learning} {Analytics} {And}
  {Knowledge}}}. \bibinfo{publisher}{ACM}, \bibinfo{address}{Indianapolis
  Indiana USA}, \bibinfo{pages}{269--270}.
\newblock
\showISBNx{978-1-4503-2664-3}
\urldef\tempurl%
\url{https://doi.org/10.1145/2567574.2567594}
\showDOI{\tempurl}


\bibitem[Zanjani(2017)]%
        {zanjani_important_2017}
\bibfield{author}{\bibinfo{person}{Nastaran Zanjani}.}
  \bibinfo{year}{2017}\natexlab{}.
\newblock \showarticletitle{The important elements of {LMS} design that affect
  user engagement with e-learning tools within {LMSs} in the higher education
  sector}.
\newblock \bibinfo{journal}{\emph{Australasian Journal of Educational
  Technology}} \bibinfo{volume}{33}, \bibinfo{number}{1} (\bibinfo{date}{April}
  \bibinfo{year}{2017}).
\newblock
\showISSN{1449-5554}
\urldef\tempurl%
\url{https://doi.org/10.14742/ajet.2938}
\showDOI{\tempurl}
\newblock
\shownote{Number: 1}.


\bibitem[Zawacki-Richter et~al\mbox{.}(2019)]%
        {zawacki-richter_systematic_2019}
\bibfield{author}{\bibinfo{person}{Olaf Zawacki-Richter},
  \bibinfo{person}{Victoria~I. Marín}, \bibinfo{person}{Melissa Bond}, {and}
  \bibinfo{person}{Franziska Gouverneur}.} \bibinfo{year}{2019}\natexlab{}.
\newblock \showarticletitle{Systematic review of research on artificial
  intelligence applications in higher education – where are the educators?}
\newblock \bibinfo{journal}{\emph{International Journal of Educational
  Technology in Higher Education}} \bibinfo{volume}{16}, \bibinfo{number}{1}
  (\bibinfo{date}{Dec.} \bibinfo{year}{2019}), \bibinfo{pages}{39}.
\newblock
\showISSN{2365-9440}
\urldef\tempurl%
\url{https://doi.org/10.1186/s41239-019-0171-0}
\showDOI{\tempurl}


\bibitem[Zeineddine et~al\mbox{.}(2019)]%
        {zeineddine_auto-generated_2019}
\bibfield{author}{\bibinfo{person}{Hassan Zeineddine}, \bibinfo{person}{Udo
  Braendle}, {and} \bibinfo{person}{Assaad Farah}.}
  \bibinfo{year}{2019}\natexlab{}.
\newblock \showarticletitle{Auto-generated ensemble model for predicting
  student success}. In \bibinfo{booktitle}{\emph{Proceedings of the {Second}
  {International} {Conference} on {Data} {Science}, {E}-{Learning} and
  {Information} {Systems} - {DATA} '19}}. \bibinfo{publisher}{ACM Press},
  \bibinfo{address}{Dubai, United Arab Emirates}, \bibinfo{pages}{1--4}.
\newblock
\showISBNx{978-1-4503-7284-8}
\urldef\tempurl%
\url{https://doi.org/10.1145/3368691.3368714}
\showDOI{\tempurl}


\bibitem[Zhang et~al\mbox{.}(2021)]%
        {zhang_undergraduate_2021}
\bibfield{author}{\bibinfo{person}{Yupei Zhang}, \bibinfo{person}{Rui An},
  \bibinfo{person}{Jiaqi Cui}, {and} \bibinfo{person}{Xuequn Shang}.}
  \bibinfo{year}{2021}\natexlab{}.
\newblock \showarticletitle{Undergraduate {Grade} {Prediction} in {Chinese}
  {Higher} {Education} {Using} {Convolutional} {Neural} {Networks}}. In
  \bibinfo{booktitle}{\emph{{LAK21}: 11th {International} {Learning}
  {Analytics} and {Knowledge} {Conference}}}. \bibinfo{publisher}{ACM},
  \bibinfo{address}{Irvine CA USA}, \bibinfo{pages}{462--468}.
\newblock
\showISBNx{978-1-4503-8935-8}
\urldef\tempurl%
\url{https://doi.org/10.1145/3448139.3448184}
\showDOI{\tempurl}


\bibitem[Zhou et~al\mbox{.}(2021)]%
        {zhou_analysis_2021}
\bibfield{author}{\bibinfo{person}{Qing Zhou}, \bibinfo{person}{Qiang Zhang},
  {and} \bibinfo{person}{Hao Li}.} \bibinfo{year}{2021}\natexlab{}.
\newblock \showarticletitle{Analysis on the level of higher universities in
  different countries using entropy weight method and analytic hierarchy
  process}. In \bibinfo{booktitle}{\emph{2021 4th {International} {Conference}
  on {Information} {Systems} and {Computer} {Aided} {Education}}}.
  \bibinfo{publisher}{ACM}, \bibinfo{address}{Dalian China},
  \bibinfo{pages}{2722--2727}.
\newblock
\showISBNx{978-1-4503-9025-5}
\urldef\tempurl%
\url{https://doi.org/10.1145/3482632.3487502}
\showDOI{\tempurl}


\end{thebibliography}

\clearpage
\onecolumn
\appendix
\section{Code Sheet}
\begin{table}[b]
\resizebox{7in}{!}{%
\begin{tabular}{lllll}
\textbf{Reference} & \textbf{Input Data} & \textbf{Target Output} & \textbf{Computational Methods} & \textbf{Study Design} \\
\cite{aguiar_engagement_2014}& Demographic; Grade/GPA; Protected Class & Retention & ML &  \\
\cite{ajoodha_forecasting_2020}& Demographic; Grade/GPA; Protected Class & Retention & ML &  \\
\cite{alghamdi_machine_2020}& Grade/GPA & Admissions & ML &  \\
\cite{altaf_student_2019}& LMS/Engagement; Grade/GPA & Retention & DL &  \\
\cite{amballoor_technological_2021}& Institutional & Institutional Planning & Statistical Methods &  \\
\cite{ameri_survival_2016}& Demographic; Grade/GPA; Enrollment/Pathways; Protected Class   & Retention & Statistical Methods &  \\
\cite{baranyi_interpretable_2020}& Demographic; Grade/GPA; Enrollment/Pathways; Protected Class & Retention & ML; DL &  \\
\cite{barber_course_2012}& Demographic; LMS/Engagement; Grade/GPA; Protected Class & Retention & Statistical Methods &  \\
\cite{benablo_higher_2018}& Demographic; Grade/GPA; Protected Class & Grade Prediction & ML &  \\
\cite{borrella_predict_2019}& LMS/Engagement; Grade/GPA; Enrollment/Pathways & Retention & ML & Theoretical Design \\
\cite{bos_student_2016}& LMS/Engagement; Student Survey & Grade Prediction & Statistical Methods &  \\
\cite{buerck_predicting_2013}& LMS/Engagement; Grade/GPA & Grade Prediction & Statistical Methods &  \\
\cite{cabrera_data_2020}& Demographic; Protected Class & Student Services & Statistical Methods &  \\
\cite{chen_pathways_2022}& Enrollment/Pathways & Pathway Advising & NLP & Theoretical Design, Participatory Design \\
\cite{chockkalingam_which_2021}& Grade/GPA; Enrollment/Pathways & Institutional Planning & ML; DL; NLP & \\
\cite{chong_data_2020}& Demographic; Grade/GPA; Protected Class & Grade Prediction & ML; DL &  \\
\cite{daud_predicting_2017}& Demographic; Protected Class & Retention & ML &  \\
\cite{dawson_prediction_2017}& Demographic; Protected Class & Retention & Statistical Methods &  \\
\cite{echeverria_presentation_2014}& - & Assessment & ML &  \\
\cite{edwards_using_2017}& LMS/Engagement; Grade/GPA; Student Survey & Grade Prediction & Statistical Methods &  \\
\cite{errahmouni_barkam_testing_2022}& Demographic; Grade/GPA; Protected Class & Retention & ML &  \\
\cite{feild_generalized_2018}& Grade/GPA & Engagement & ML &  \\
\cite{fiorini_application_2018}& Grade/GPA; Enrollment/Pathways & Retention & ML; & Participatory Design \\
\cite{gamie_layered-analysis_2019}& Demographic; LMS/Engagement; Grade/GPA; Enrollment/Pathways;   Protected Class & Engagement & ML &  \\
\cite{gatbonton_employability_2018}& Grade/GPA & Institutional Planning & Rules-Based &  \\
\cite{hien_intelligent_2018}& Student Survey & Student Services & DL; NLP &  \\
\cite{hlosta_ouroboros_2017}& Demographic; LMS/Engagement; Protected Class & Retention & ML &  \\
\cite{jayaprakash_open_2014}& Demographic; LMS/Engagement; Grade/GPA; Protected Class & Retention & ML &  \\
\cite{jiang_time_2019}& LMS/Engagement; Grade/GPA; Enrollment/Pathways & Pathway Advising & DL &  \\
\cite{jiang_towards_2021}& Demographic; Grade/GPA; Enrollment/Pathways; Protected Class & Grade Prediction & DL & Speculative Design \\
\cite{kaur_causal_2019}& Grade/GPA & Pathway Advising & Statistical Methods &  \\
\cite{kostopoulos_estimating_2015}& Demographic; Protected Class & Retention & ML &  \\
\cite{kung_interpretable_2020}& Demographic; LMS/Engagement; Grade/GPA; Protected Class & Grade Prediction & ML & Speculative Design \\
\cite{lang_forecasting_2021}& Enrollment/Pathways & Pathway Advising & ML &  \\
\cite{luzardo_estimation_2014}& - & Assessment & ML &  \\
\cite{ma_using_2018}& Demographic; Grade/GPA; Protected Class & Grade Prediction & ML; DL &  \\
\cite{maramag_assessing_2019}& Demographic; Student Survey; Protected Class & Retention & Statistical Methods &  \\
\cite{muslim_rule-based_2016}& LMS/Engagement & Engagement & Rules-Based &  \\
\cite{nespereira_is_2014}& LMS/Engagement & Grade Prediction & Statistical Methods; ML &  \\
\cite{obeid_ontology-based_2018}& Student Survey & Pathway Advising & ML &  \\
\cite{ojajuni_predicting_2021}& Demographic; Grade/GPA; Protected Class & Grade Prediction & ML &  \\
\cite{oreshin_implementing_2020}& Demographic; Protected Class & Retention & ML &  \\
\cite{orong_predictive_2020}& Demographic; Protected Class & Retention & ML &  \\
\cite{pardos_data-assistive_2019}& Enrollment/Pathways & Pathway Advising & ML; NLP &  \\
\cite{pardos_designing_2020}& Enrollment/Pathways & Pathway Advising & DL; NLP & Participatory Design \\
\cite{perez-sanagustin_can_2021}& LMS/Engagement & Retention & ML; &  \\
\cite{pokorny_machine_2021}& Institutional; Enrollment/Pathways & Institutional Planning & ML &  \\
\cite{rijati_multi-attribute_2018}& Demographic; Grade/GPA & Student Services & ML &  \\
\cite{rogers_modest_2014}& Demographic; LMS/Engagement; Grade/GPA; Enrollment/Pathways; Protected Class & Grade Prediction & Statistical Methods &  \\
\cite{sallaberry_comparison_2021}& - & Assessment & ML; DL &  \\
\cite{sekeroglu_student_2019}& Grade/GPA & Grade Prediction & ML; DL &  \\
\cite{soleimani_comparative_2021}& LMS/Engagement & Engagement & ML; &  \\
\cite{staudaher_predicting_2020}& Demographic; Protected Class & Admissions & ML &  \\
\cite{trandafili_discovery_2012}& Grade/GPA; Enrollment/Pathways & Grade Prediction & ML; &  \\
\cite{van_goidsenhoven_predicting_2020}& LMS/Engagement & Grade Prediction & ML &  \\
\cite{wang_first-gen_2022}& Student Survey & Student Services & ML; DL &  \\
\cite{wham_forecasting_2017}& Demographic; Grade/GPA; Enrollment/Pathways; Protected Class & Grade Prediction & ML &  \\
\cite{yu_should_2021}& Demographic; Grade/GPA; Enrollment/Pathways; Protected Class & Retention & ML & Speculative Design \\
\cite{yu_educational_2014}& LMS/Engagement & Grade Prediction & Statistical Methods &  \\
\cite{zeineddine_auto-generated_2019}& Demographic; Protected Class & Grade Prediction & ML; DL &  \\
\cite{zhang_undergraduate_2021}& Demographic; Grade/GPA; Protected Class & Grade Prediction & DL &  \\
\cite{zhou_analysis_2021}& Institutional; Enrollment/Pathways & Institutional Planning & ML; &  \\
\end{tabular}%
}
\vspace{9pc}\end{table}

\end{document}